\documentclass[aip,reprint]{revtex4-1}

\usepackage{graphicx,color}
\usepackage{url}

\begin{document}


\title{An ultra-high vacuum scanning tunneling microscope operating at sub-Kelvin temperatures and high magnetic fields for spin-resolved measurements} 



\author{C. Salazar}
\email[]{c.d.salazar.enriquez@ifw-dresden.de}
\affiliation{Leibniz Institute for Solid State and Materials Research, IFW-Dresden, Helmholtzstra{\ss}e 20, 01069 Dresden, Germany}

\author{D. Baumann}
\email[]{d.baumann@ifw-dresden.de}
\affiliation{Leibniz Institute for Solid State and Materials Research, IFW-Dresden, Helmholtzstra{\ss}e 20, 01069 Dresden, Germany}

\author{T. H{\"a}nke}
\affiliation{Leibniz Institute for Solid State and Materials Research, IFW-Dresden, Helmholtzstra{\ss}e 20, 01069 Dresden, Germany}

\author{M. Scheffler}
\affiliation{Leibniz Institute for Solid State and Materials Research, IFW-Dresden, Helmholtzstra{\ss}e 20, 01069 Dresden, Germany}

\author{T. K{\"u}hne}
\affiliation{Leibniz Institute for Solid State and Materials Research, IFW-Dresden, Helmholtzstra{\ss}e 20, 01069 Dresden, Germany}

\author{M. Kaiser}
\affiliation{Leibniz Institute for Solid State and Materials Research, IFW-Dresden, Helmholtzstra{\ss}e 20, 01069 Dresden, Germany}

\author{R. Voigtl{\"a}nder}
\affiliation{Leibniz Institute for Solid State and Materials Research, IFW-Dresden, Helmholtzstra{\ss}e 20, 01069 Dresden, Germany}

\author{D. Lindackers}
\affiliation{Leibniz Institute for Solid State and Materials Research, IFW-Dresden, Helmholtzstra{\ss}e 20, 01069 Dresden, Germany}

\author{B. B{\"u}chner}
\affiliation{Leibniz Institute for Solid State and Materials Research, IFW-Dresden, Helmholtzstra{\ss}e 20, 01069 Dresden, Germany}
\affiliation{Center for Transport and Devices, Technische Universit{\"a}t Dresden, 01069 Dresden,Germany}
\affiliation{Institute of Solid State Physics, Technische Universit{\"a}t Dresden, 01069 Dresden,Germany}

\author{C. Hess}
\email[]{c.hess@ifw-dresden.de}
\affiliation{Leibniz Institute for Solid State and Materials Research, IFW-Dresden, Helmholtzstra{\ss}e 20, 01069 Dresden, Germany}
\affiliation{Center for Transport and Devices, Technische Universit{\"a}t Dresden, 01069 Dresden,Germany}

\date{\today}

\begin{abstract}
We present the construction and performance of an ultra-low temperature scanning tunneling microscope (STM), working in ultra-high vacuum conditions (UHV) and in high magnetic fields up to 9 T. The cryogenic environment of the STM is generated by a single shot $^3$He magnet cryostat in combination with a $^4$He dewar system. At base temperature (300~mK), the cryostat has an operation time of approximately 80 hours. The special design of the microscope allows the transfer of the STM head from the cryostat to a UHV-chamber system, where samples and STM-tips can be easily exchanged. The UHV chambers are equipped with specific surface science treatment tools for the functionalization of samples and tips, including high-temperature treatments and thin film deposition. This, particularly, enables spin-resolved tunneling measurements. We present test measurements using well known samples and tips based on superconductor and metallic materials such as LiFeAs, Nb, Fe and W, respectively. The measurements demonstrate the outstanding performance of the STM with high spatial and energy resolution as well as the spin-resolved capability.
\end{abstract}
             

\maketitle

\section{Introduction}
Current research on condensed matter physics focuses on the study of correlated electron systems, that exhibit interesting electronic collective effects measurable mostly at small energy scales (e.g. meV-$\mu$eV). In this regard, scanning tunneling microscopy and spectroscopy (STM/STS) operating at cryogenic temperatures has played a relevant role due to its high energy and spatial resolution. Prominent examples are the spectroscopic maps of electronic wave functions in metals, topological insulators, unconventional superconductors, among others\cite{crommie1993confinement,Yazdani1997,Hoffman2002,yee2015spin, fu2014imaging, kohsaka2012visualization, hanke2012probing}. Furthermore, spectroscopic maps at low temperature have strongly contributed to the establishment of the so called spin-polarized scanning tunneling microscopy technique (SP-STM)\cite{bode2003spin,wiesendanger2009spin}. However, experiments combining ultra-low temperatures (specifically sub-Kelvin) and spin sensitivity are still rare and thus highly desired. 
Different STMs operating at sub-Kelvin temperatures have been built, where some of them are rely on $^3$He refrigerators, Joule-Thompson refrigerators and others on dilution refrigerators\cite{kamlapure2013350, wiebe2004300, assig201310, roychowdhury201430, galvis2015three, singh2013construction, song2010invited, Yazdani2013rsi, Wulfhekel2011rsi, Morgenstern2017rsi, unisoku, specs}. SP-STM, on the other hand typically requires the combination of the STM with ultra-high vacuum (UHV) environments.\cite{bode2003spin,wiesendanger2009spin}

A bunch of interesting experiments in the field of condensed matter at extremely low temperatures and with spin sensitivity might provide not only new insight into the physics of exotic materials but also potentially contribute to the design of novel applications. 
Some topics which might profit from such experiments are: i) unconventional superconductivity in correlated materials such as heavy fermion, iron pnictide and cuprate compounds, ii) $p$-wave superconductivity, iii) quantum Hall effect, iv) topologically non-trivial systems such as skyrmion systems, topological insulators or Weyl semimetals\cite{Scalapino2012,Davis2013,jourdan1999superconductivity,ishida1998spin,tsui1982two,1367-2630-18-6-065003,Feldman316,PhysRevLett.107.127205}. 

Here we report the construction of an STM system operating at temperatures about 300~mK\cite{Baumann2011}. The low-temperature environment is realized by a $^3$He evaporation cryostat. The cryostat is coupled to a UHV environment of three UHV chambers where standard preparation and analysis surface tools are installed in order to allow the in-situ preparation of samples and tips for spin-sensitive measurements. The system additionally is equipped with a superconducting magnet which allows to apply an external magnetic field (up to 9~T) perpendicular to the sample surface.
We report various characterizing measurements on well-known metallic magnetic and superconducting samples and different, partially magnetized tips in order to demonstrate the performance of the microscope\cite{Salazar2016, Scheffler2015}.  
                
\section{Instrument and design}

The instrumental design attempts to reduce as much as possible mechanical and electromagnetic noise that might interfere in the energy resolution determined by the thermal limit of the system. In order to reduce the mechanical noise, various aspects and components have been considered, in particular: i) the STM host and auxiliary rooms, ii) a pneumatic damping system, iii) a  metallic host frame, and iv) the rigidity of the STM head. On the other hand, electromagnetic noise has been reduced by using noise filters and avoiding multiple grounds.

\begin{figure*}
\includegraphics[scale=0.65]{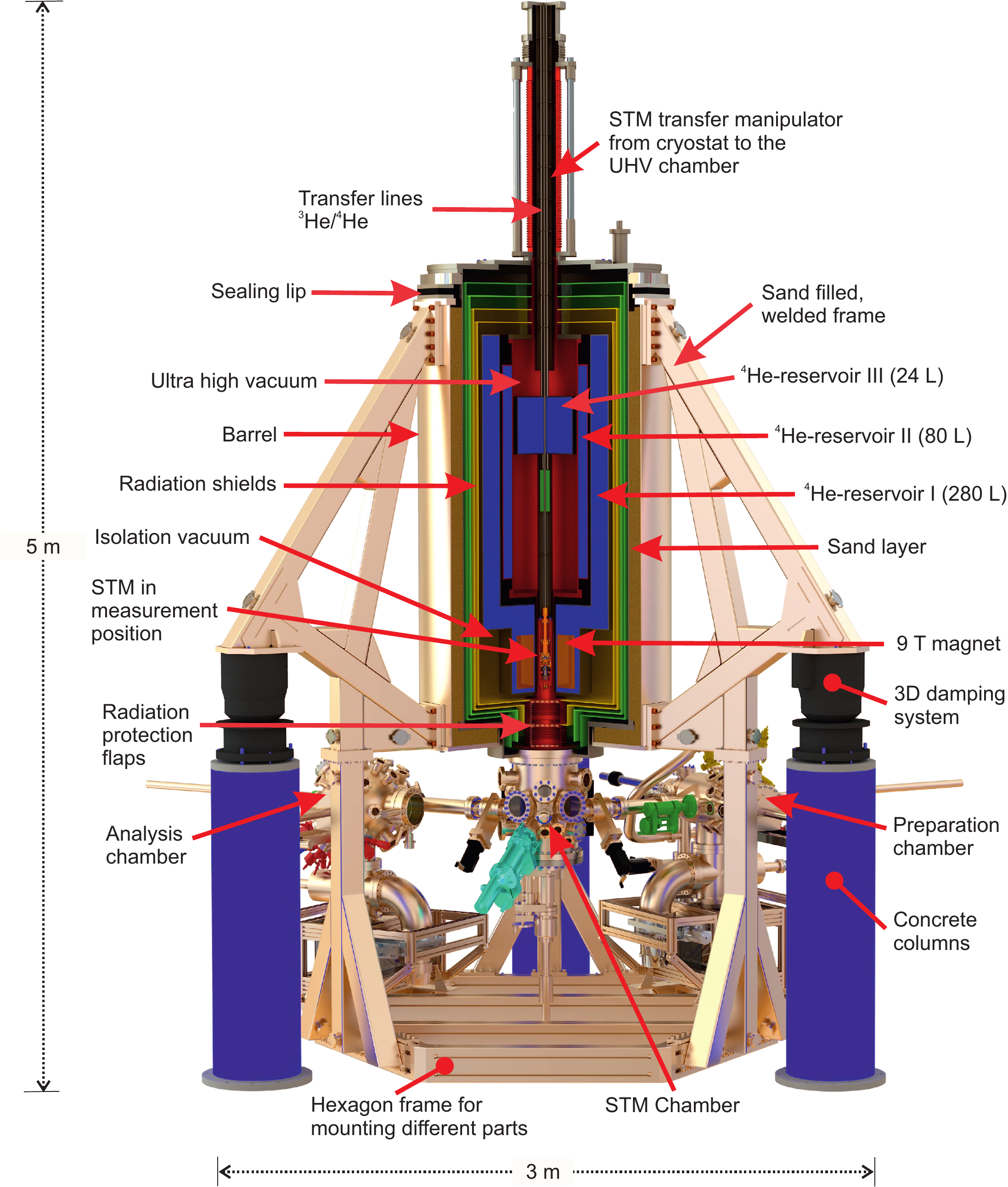}
\caption{Overview of the STM system with cryostat, damping system and UHV chambers. The section view of the cryostat shows the He-reservoirs, the 9T magnet as well as the STM in the measurement position.}
\label{fig:overview}
\end{figure*}

FIG.~\ref{fig:overview} shows an overview of the STM system. The most prominent part is the cryostat system which consists of a combination of several $^4$He baths  and a single shot $^3$He magnet cryostat. In the inner part of the cryostat a superconducting magnet and a rod-manipulator which hosts the STM head are installed. The cryostat is held in a barrel that is part of a metallic frame, which also hosts three UHV chambers used for the preparation of samples and tips. The complete system rests on a pneumatic damping system consisting of three equally spaced columns (see schematic in FIG.~\ref{fig:overview}). More details will be described in the following sections.

\subsection{Mechanical vibration isolation}

The STM system requires high mechanical stability in order to achieve high spatial resolution. The decoupling mechanism of the STM from the building can be described starting with the decoupling of the cryostat. The cryostat is suspended at the top of a metallic barrel through the assistance of an elastic, electrically insulating sealing lip (see FIG.~\ref{fig:overview}). In this way a direct mechanical and electrical contact between the metallic barrel and the cryostat is prevented. Furthermore, the damping elements avoid the direct transmission of remaining mechanical noise from the barrel body to the cryostat. 

The barrel is filled with fine grained dried sand, which screens the cryostat of sounds coming from the surroundings. The barrel itself is part of a metallic frame that connects to a hexagonal structure in the lowest part of the system. The hollow spaces of the vertical frame which supports the hexagonal structure in the frame is also filled with sand. The hexagonal structure provides stability to the whole system and hosts  two of the three UHV chambers (i.e. the preparation and analysis chambers). The third chamber (labeled STM chamber) is directly bolted to the cryostat and has no rigid mechanical connection to the other two UHV chambers, where the UHV connection is realized through edge-welded bellows. The entire metallic frame sits on three damping elements equally distributed forming 120$^{\circ}$ between them. Each damping element has a horizontal (BILZ HAB 24000) and a vertical (BILZ BIAIR 2.5-ED/HD) air-pillow with an air pressure regulator \cite{bilz}. Thus, the entire STM system is decoupled from all three possible vibration directions of the building. 

Each air damping unit sits on a column made of concrete. The height of the columns were chosen such that the midpoint of any tilting vibration of the structure, lies exactly at the measurement position of the STM. Therefore, the design attempts that vibrational and rotational modes of the system have low impact on the measurement.
                     
\subsection{Cryogenics and magnet}

The bottom loading $^3$He cryostat system (model HE-3-BLSUHV-STM) which provides the low-temperature vacuum environment and the magnetic field has been manufactured by Janis Research Company\cite{janis}. The base temperature of about 300~mK of the system is provided by a single-shot $^3$He evaporation cryostat in combination with a low-noise 1K-pot fed by liquid $^4$He and several $^4$He baths.       

\subsubsection{Heat isolation}

The first part of the system, from outside to inside, which  protects the STM from external heat radiation are the radiation shields. These super-isolated shields are exposed to the exhaust steam of $^4$He reservoirs~I and II, reaching typical temperatures in the range from 65 K at the most internal shield (yellow color in cryostat schematic) to 240 K at the most outer shield (green color in cryostat schematic).

At the junction between the cryostat and the UHV STM chamber, three radiation flaps avoid the entrance of room temperature heat radiation. These flaps can be pushed open by the STM-1K-shield (see below) when the STM head unit is moved from the measurement position, located at the center of the magnetic field (see schematic in FIG.~\ref{fig:overview}), towards the UHV chamber. If the STM head unit is back to the measurement position, the three flaps close automatically through spring forces. The most internal plate is additionally connected to the $^4$He reservoir~I through a thermal copper connection, reaching temperatures less than 5.5 K. It guarantees a good protection of the $^4$He reservoirs of external heat radiation.

There are two regions in the cryostat of isolation vacuum: an external one (dark yellow in the cryostat), which reaches pressure less than 10$^{-6}$ mbar and that encloses $^4$He reservoirs I and II, and an internal one, at the center of the cryostat (red in the cryostat), which reaches pressure less than 10$^{-10}$ mbar and encloses $^4$He reservoir III.      

\subsubsection{$^4$Helium reservoirs}

Reservoir I has a capacity of 280 liters and is the largest in the system. It serves to cool down the core area of the cryostat and to protect it from external heat radiation. Furthermore, reservoir I hosts the superconducting magnet, which can provide vertical magnetic fields up to 9~T (i.e. perpendicular to the sample surface). Additionally, a connection between reservoirs~I and II refrigerates the UHV-tube, where the STM is placed in the measurement position.  

Reservoir II has a capacity of 80 liters and is concentrically located inside of reservoir~I. It is located between an isolation vacuum layer and the core UHV area of the cryostat. This reservoir protects reservoir III and the $^3$He stage that are inside the UHV area against thermal radiation. Therefore, reservoir II should all the time be operated with more than 75\% of its helium capacity in order to guarantee a thermal radiation to the $^3$He stage of about 4.2~K.           

Reservoir III has a capacity of 24 liters and is located in the UHV area of the cryostat. The reservoir is attached to a vertical manipulator where the $^3$He stage and the STM are also attached. The purpose of this reservoir is to supply liquid $^4$He to the 1~K pot of the $^3$He cryostat.

If the 1 K-pot is used, the helium consumption for the reservoir I is about 4.7\% per day with a standing time of about 21 days, for reservoir II is 6.3\% per day with a standing time of about 15 days and for reservoir III is 18,2\% per day with a standing time of about 5 days. If the 1 K-pot is not used, the helium consumption for the reservoir I is about 4.4\% per day with a standing time of 22 days, for reservoir II is 6.6\% per day with a standing time of about 15 days and for reservoir III is 11,5\% per day with a standing time of about 8 days.     

\subsubsection{$^3$Helium cryostat}

\begin{figure}
\includegraphics[scale=0.38]{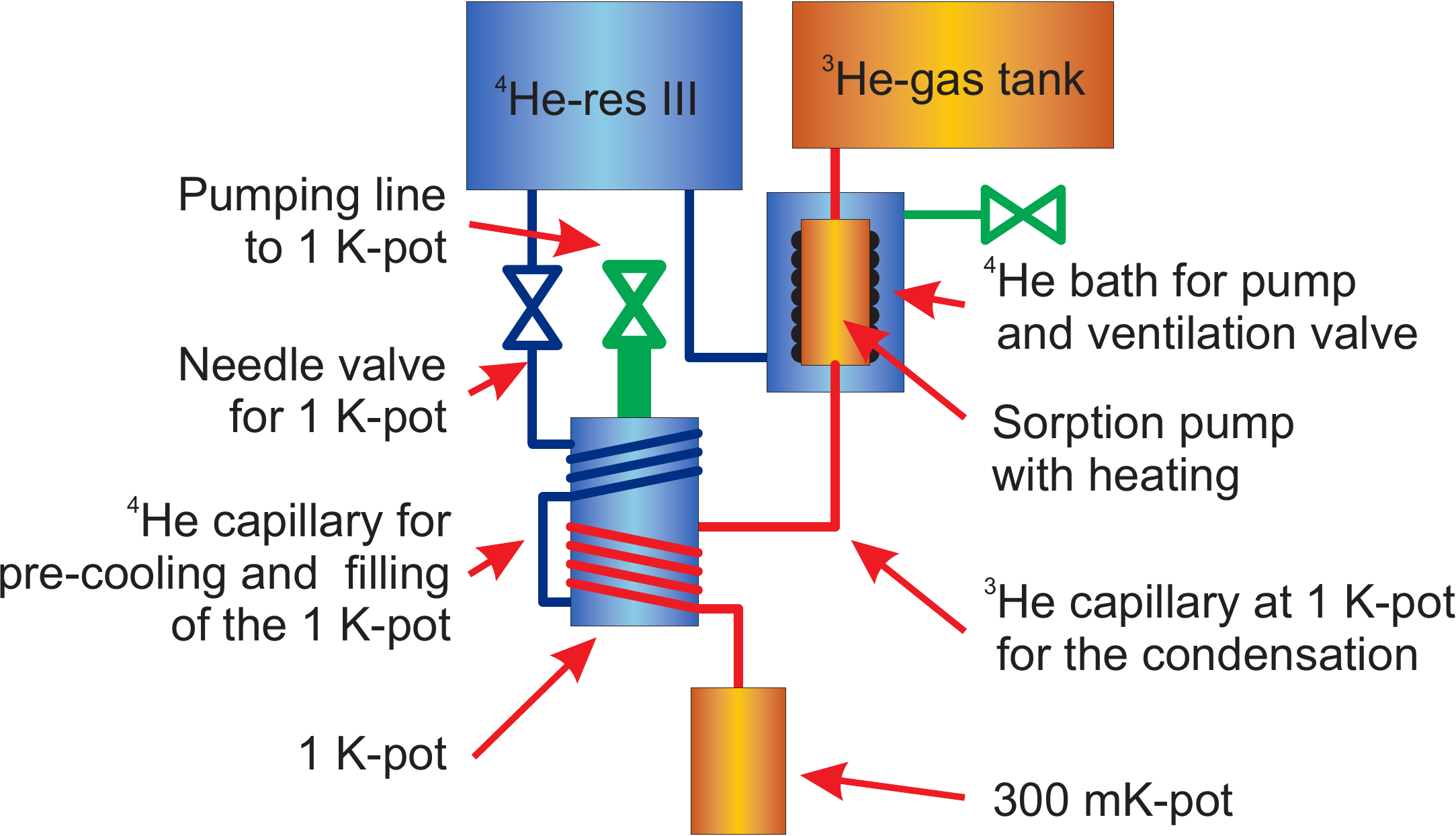}
\caption{Functional diagram of the $^3$He stage of the cryostat showing the open $^4$He circuit and the closed $^3$He circuit. Reproduced from Reference 30.}
\label{fig:3He-schematics}
\end{figure}

The $^3$He cryostat operates with two separated helium circuits (see FIG.~\ref{fig:3He-schematics}). One circuit employs liquid $^4$He provided by reservoir III, whereas the second circuit is a closed $^3$He-circuit. The purpose of the $^4$He-circuit is to supply the 1 K-pot, i.e. a container with a volume of approximately 65 ml, with liquid $^4$He while pumping on the gas phase of the $^4$He with a rotary pump. The temperature thus reached in the 1 K-pot is about 1.3~K, depending on the adjustment of a needle valve on the pumping line. This low temperature is required to condensate $^3$He as the first step to reach the base temperature of the system. The $^3$He condensation takes place in the external capillaries which wrap the 1 K-pot. Typically, a proper adjustment of the needle valve will let the 1 K-pot stay for approximately 5 days at about 1.3 K and additionally operate with the minimum influence to the noise of the microscope\cite {Raccanelli2001}. Afterward, reservoir~III should be filled again.

The $^3$He circuit has three parts: i) the $^3$He-gas tank, ii) an activated carbon adsorption pump and iii) the 300 mK-pot (see FIG.~\ref{fig:3He-schematics}). All components are connected through capillaries. Between the adsorption pump and the 300 mK-pot the capillaries are thermally coupled to the 1 K-pot. The $^3$He tank has a volume of about 1200 cm$^3$ where gaseous $^3$He is stored under a pressure of approximately 7~bar. In order to reach the base temperature, the process starts with the cooling of the activated carbon pump at about 4~K, so that the complete $^3$He is absorbed on the activated carbon surface. To condensate $^3$He, the activated carbon pump is first heated to about 40~K using an installed $25~\Omega$-heater. Thereby, the $^3$He is released and condensates afterward at the capillaries which are wrapped around the 1 K-pot. The liquid $^3$He subsequently drops into the 300 mK-pot and the microscope head cools down to about 2~K. After reaching this temperature, it takes around 2~hours to condense the required amount of about 26~ml of liquid $^3$He. After finishing the condensation process, the activated carbon pump should again be cooled down to 4~K. This starts a cryo-pumping effect which reduces the vapor pressure and the boiling temperature of the $^3$He by absorbing the gaseous $^3$He on the activated carbon surface. This process, in approximately 10 minutes, leads to a temperature of about 300~mK at the 300 mK-pot and the STM-head which can be maintained for approximately 80 hours until all liquid $^3$He is fully evaporated.                  

\subsubsection{Temperature variability}

The system allows to vary the temperature from 300~mK up to 40~K while maintaining high stability required for STM measurements. Between 300 mK and the boiling point of $^3$He ($T = 3.2$~K), the temperature can be adjusted by changing the pump rate at the 300 mK-pot. Technically, this requires to heat the sorption pump up through the $25~\Omega$-heater. Higher temperatures up to about 40~K can be reached by actively heating the 300 mK-pot with a second $25~\Omega$-heater and controlling the temperature using a CERNOX and a RuO temperature sensor\cite{lakeshore}.        

\subsection{STM head and electronics}

\begin{figure}
\includegraphics[scale=0.38]{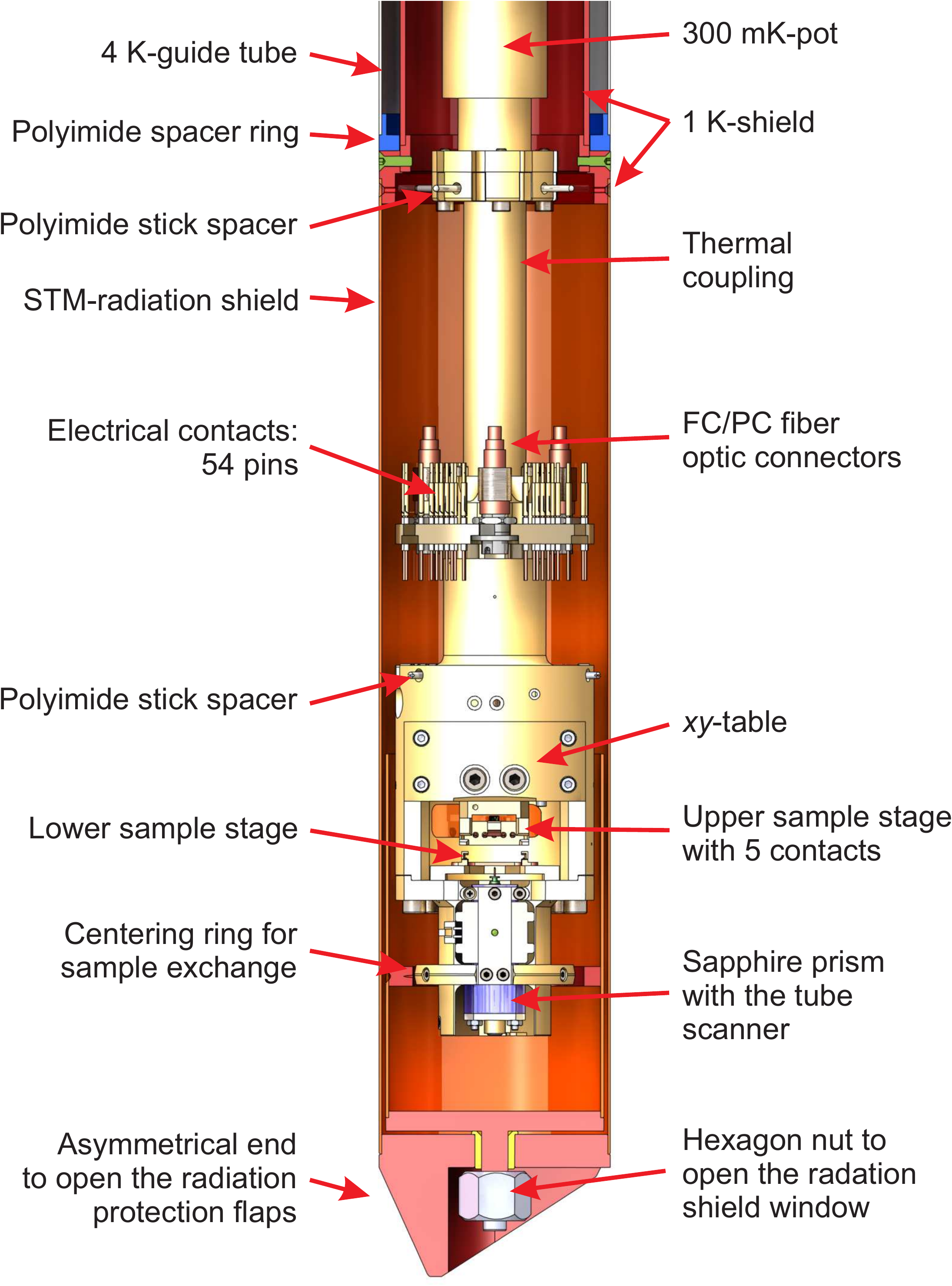}
\caption{STM with the thermal coupling and the 300 mK-pot. The inner and outer 1 K-shield are cut with the purpose to see the STM. Dimensions of the 1 K-shield: 340 mm of length and 70 mm of diameter. Reproduced from Reference 30.}
\label{fig:STM-head}
\end{figure}

The central STM unit of the system is shown in FIG.~\ref{fig:STM-head}. The STM design is based on previously reported ones \cite{haude2001, wachowiak2003, hanke2005,Schlegel2014}. The materials employed in the STM construction satisfy conditions such as high mechanical firmness, good thermal conduction, similar thermal expansion coefficients, and weak magnetic response of the used materials. Last but not least, the materials were chosen to be easily machinable. 

The chosen materials and the STM unit's design support high stability in the tunneling junction and ensures versatility for non conventional transport and tunneling experiments. As for the latter, the design includes the following features: i) samples and tips can be exchanged inside the UHV, ii) the sample can be moved relatively with respect to the tip through a coarse positioning mechanism (so-called $xy$-table), and iii) the STM sample holder receptacle is equipped with electrical connection to allow four-point transport measurements, including a gate-electrode, simultaneously to the tunneling experiment. The STM unit is surrounded by a protection shield cooled to 1 K (so-called 1 K-shield) in order to avoid direct radiative heat intake from warmer parts of the instrument. The 1 K-shield possesses a window which can be opened and closed with the purpose to have access to the STM through an external manipulator in the transfer position (not shown in FIG. ~\ref{fig:STM-head}) and to ensure a homogeneous thermal radiation environment.

\subsubsection{Thermal coupling and 1 K-shield}

The thermal coupling connects the actual STM unit with the 300~mK pot of the cryostat (see FIG.~\ref{fig:STM-head}) and is made of oxygen free high thermal conductivity copper. In order to achieve an excellent thermal connection, the contact surfaces were polished, and subsequently the complete piece has been gold-coated with a thickness of approximately 5~$\mu$m. The thermal coupling is fixed via four (DIN 912-M3x25 mm) titanium screws at the side of the 300~mK-pot and with three (DIN 912-M3x25 mm) titanium screws on the STM side. In this way, high contact pressure in both flanges is ensured. 
The 300 mK-pot is suspended by capillaries with a length of about 300 mm with the purpose to guide the $^3$He to and from the pot, ensuring low heat input from the 1 K-pot. Moreover, four polyimide sticks ensure that the 1 K-shield keeps the central position with respect to the thermal coupling device. The form of the flange of the 300 mK-pot lets the cables and optic fibers (for optional future combination with optical experiments) pass through the inner diameter of the thermal coupling device and avoids the contact with the 1 K-shield. 

The 1 K-shield has a compact connection to the 1 K-pot and is made of oxygen-free copper suitable for UHV. It has a length of about 340 mm and a diameter of 70 mm. The complete piece is also coated with  approximately 5~$\mu$m of gold. The 1 K-shield has an asymmetric lower tail that serves to open the radiation protection flaps which separate the STM UHV-chamber from the cryostat. Furthermore, in the 1 K-shield end, there is an hexagonal insert that lets the user manipulate the 1 K-shield window, providing access to the STM through an external manipulator.

\subsubsection{STM body}                  

\begin{figure}
\includegraphics[scale=0.38]{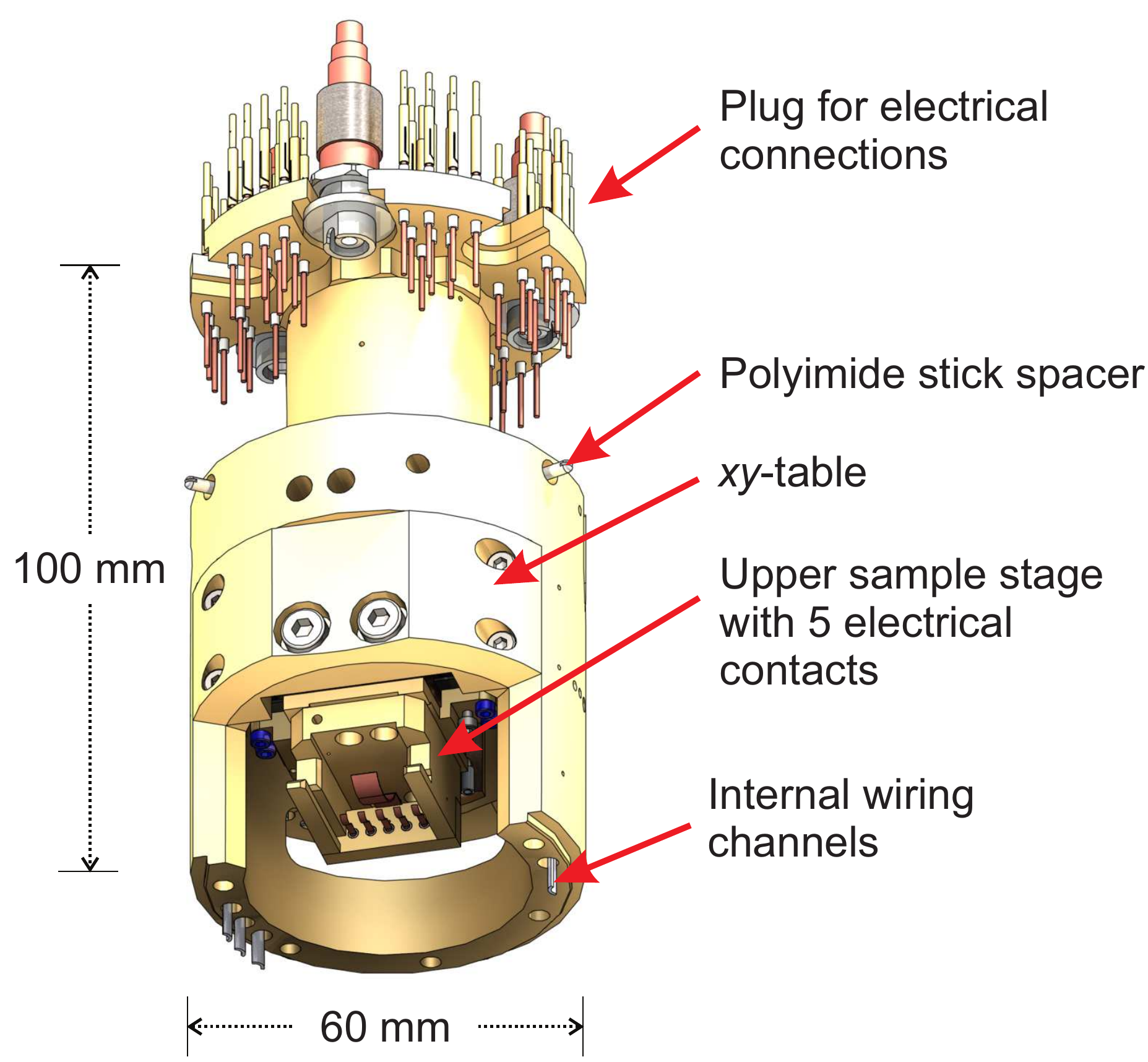}
\caption{STM-body-sample unit and electrical connection stage. Reproduced from Reference 30.}
\label{fig:sample-unit}
\end{figure}

\begin{figure}
\includegraphics[scale=0.38]{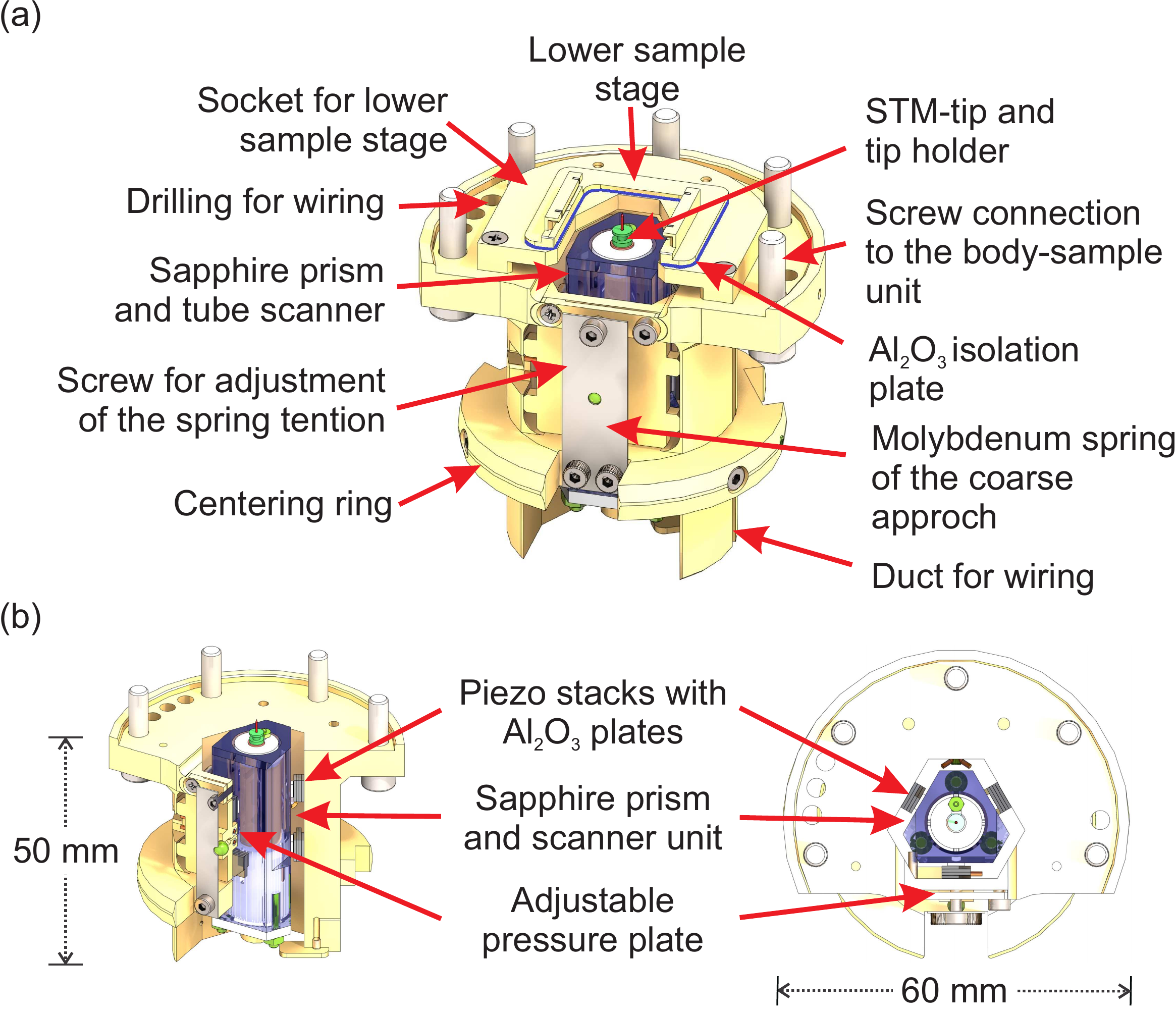}
\caption{(a) STM-body-tip unit and (b) lateral and top views of the unit. Reproduced from Reference 30.}
\label{fig:tip-unit}
\end{figure}

The STM body is made of gold-coated phosphorus bronze and can be divided into two parts: the sample unit and the tip unit (see FIG.~\ref{fig:sample-unit} and FIG. ~\ref{fig:tip-unit}). These two parts are connected with five (DIN 912-M4x20 mm) titanium screws. This massive connection supports the rigidity of the STM body and ensures a good thermal conductivity between both parts.

The sample unit consists of the upper sample receptacle part (labeled \textit{upper sample stage}),  which is directly installed in the $xy$-table. This part has five contact pins for possible additional measurements (e.g. Hall probes and temperature sensors).

The $xy$-table is a device installed in the sample-unit that enables the relative movement in the $xy$-plane of the sample with respect to the tip. This device provides the possibility to approach in different spots of the sample surface without completely warming the sample. The procedure of changing the sample-tip spot takes just few minutes until the system returns to stable conditions. The $xy$-table thus circumvents an otherwise manual manipulation of the sample position which typically requires several hours. The $xy$-table movement is  carried out via the WALKER-principle\cite{patent}, providing the required high stability. The $xy$-table is based in a crossbred sapphire prism using six piezo stacks for the movement in the $x$-axis as well as six piezo stacks for the movement in the $y$-axis.

The plug for electrical connections of the STM is located at the upper end of the sample unit. The plug contains gold-plated copper pins, which are isolated from the metal body by aluminum oxide feed-troughs and insulating glue (EpoTek H70E)\cite{epotek}. Additionally, the plug contains three FC/PC- fiber optic connectors for potential future combination with optical experiments. This connector part matches with a complementary counterpart at the thermal coupling of the cryostat. This characteristic of the STM design facilitates the separation of the STM from the cryostat in the the case of maintenance.
 
The tip unit is shown in FIG.~\ref{fig:tip-unit}. In the upper part there is a second sample receptacle (labeled \textit{lower sample stage}) which can only be used for standard STM measurements. The lower sample stage is glued with insulating glue (EpoTek H70E)\cite{epotek} to a socket, which is screwed in the STM body. Between the socket and the lower sample stage there is a 0.1 mm aluminum oxide plate that provides electric insulation and good thermal connection to the STM body. Another important part is the scanner unit that consists of the sapphire prism and the piezo tube scanner\cite{binnig1986}. The bottom part of tube scanner is mounted in an aluminum oxide socket with insulating glue and fixed with three (M1.5x3 mm) titanium screws in the sapphire prism. The top part of the tube scanner has an aluminum oxide socket, a tip stage made of molybdenum and a spring made of cooper-beryllium for the tip reception. The electrical connections of \textit{x} and \textit{y} electrodes of the tube scanner are made with thin (diameter: 0.1 mm) polyimide isolated oxygen free high thermal conductivity cooper wires. On the other hand, cooper coaxial wires (outer diameter: 0.5 mm) connect the tip (tunneling current)  and \textit{z} electrode in the tube scanner. The electrical connection of the wires was made with conducting glue (EpoTek H20E)\cite{epotek}. The scanner unit moves upwards or downwards through piezo stack actuators exploiting the slip-stick mechanism\cite{patent}. The efficiency of the coarse approach movement can be adjusted through the molybdenum spring. It is essential for the first optical and the final automatic approach of the tip to the sample.  

\subsubsection{Mechanism for tip/sample exchange}     

One important property of the STM system is the possibility to exchange the tip/sample without warming up or venting the system. A step by step procedure to transfer the tip/sample from the magnet center to the STM chamber stars with the movement of the STM transfer manipulator (see FIG.1) to the UHV STM chamber. In the chamber, the 1 K-shield window can be opened through an external manipulator that fits in the hexagon nut located at the bottom part (see FIG.3). After that, the tip/sample is reached by a wobble stick that can take the tip/sample to a sample carousel. The tip/sample exchange procedure is assisted by the coarse approach/retract mechanism of the scanner unit.   

\begin{figure}
\includegraphics[scale=0.38]{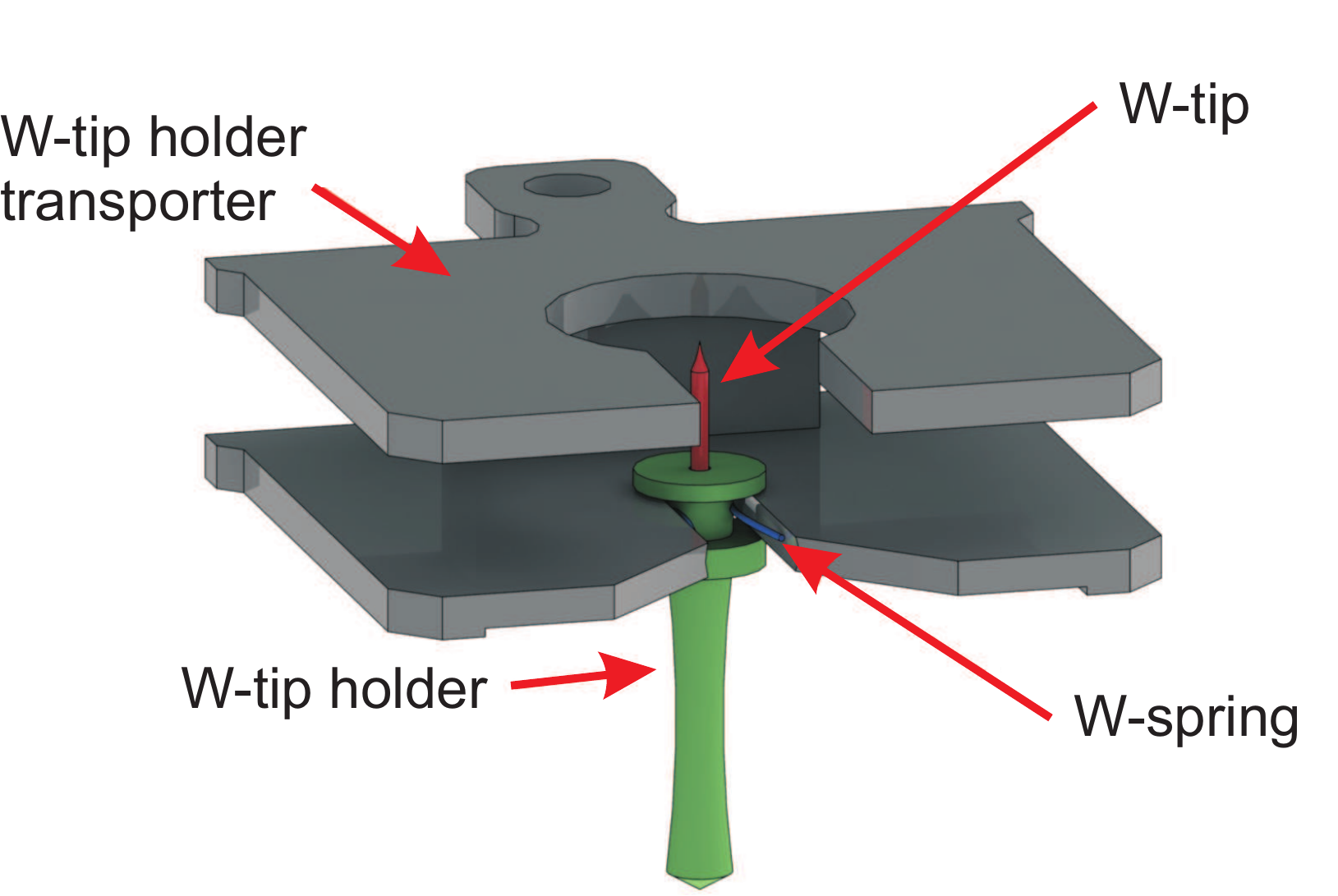}
\caption{Tip holder-transporter designed for tips exchange and treatment. Dimensions: 18 mm x 15 mm x 5 mm. Reproduced from Reference 30.}
\label{fig:tip-transporter}
\end{figure}

The tip  is transported with the assistance of a special tip holder-transporter shown in FIG.~\ref{fig:tip-transporter}. This holder-transporter is made of tungsten and has two parallel plates. The one at the bottom hosts the tip which is fixed in the tip holder. Additionally, the tip holder is secured by a tungsten spring wire. The second plate at the top protects the tip of any undesirable contact while it is moved. Furthermore, it has an opening of about 7 mm in diameter which enables the tip to be reached by electron or ion beams for cleaning purposes as well as for tip functionalization, such as coating with magnetic films required for spin-resolved measurements.

The procedure of inserting a sample from ambient conditions to the UHV system (first into the entry-lock chamber, see UHV system) takes about 1 hour. The sample transferring from the entry-lock chamber to the STM-measuring position takes about 1 h. Once the sample is in the measuring position, it takes about 3 hours to carry out the corresponding cooling and thermalization processes in order to start a STM measurement at base temperature.         

\subsubsection{Electronic connections of the STM system}
 
\begin{figure}
\includegraphics[scale=0.4]{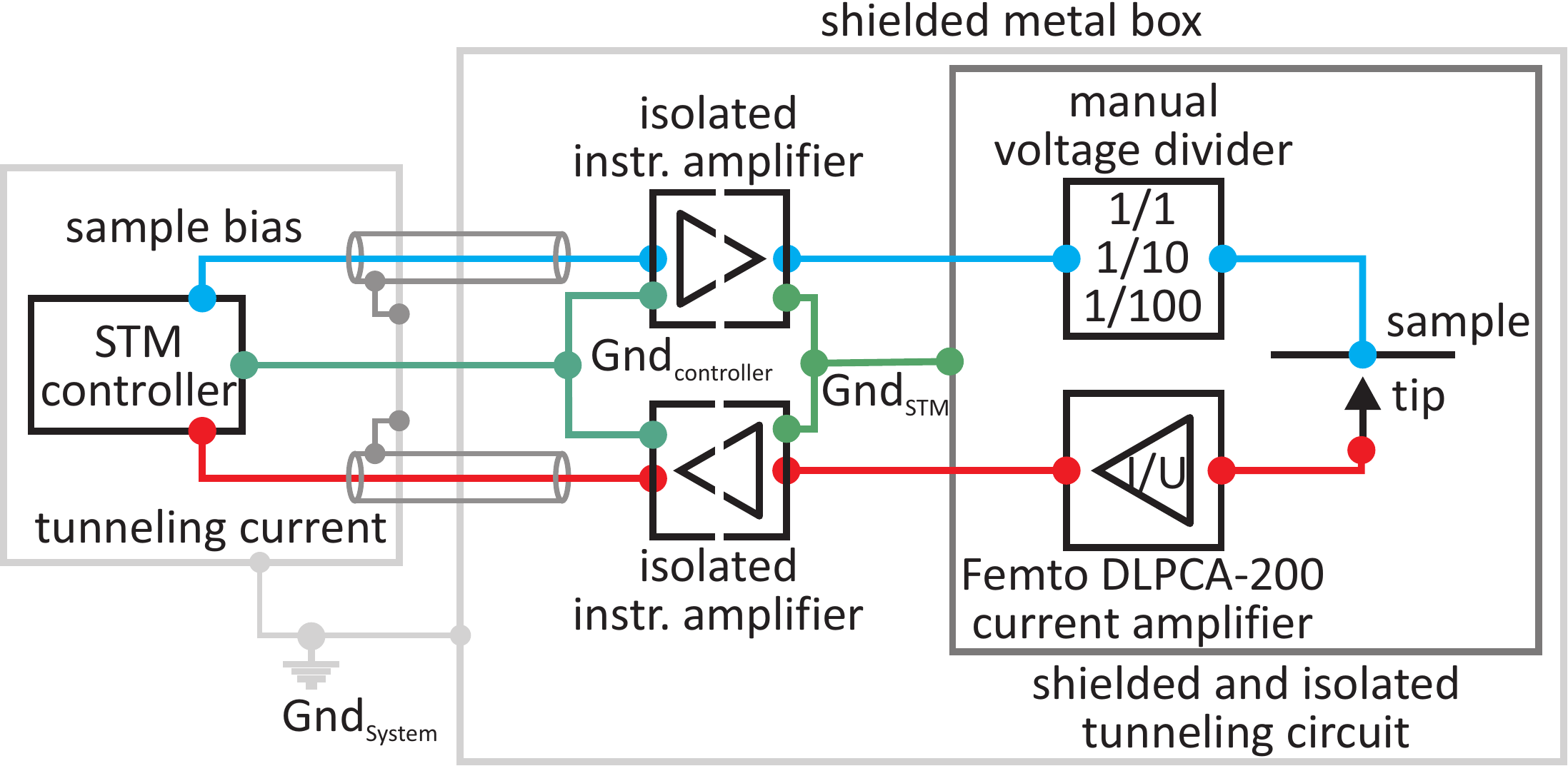}
\caption{Electronic connections of the STM system. The amplifier of the current loop is in a RF-screened box mounted and only connected to the STM controller. In order to prevent occurring ground loops, isolated voltage amplifiers with an amplifier factor of 1 are used. Inside the box, all signal lines from the sample / tip to the voltage divider / amplifier are surrounded by a closed screen. Reproduced from Reference 30.}
\label{fig:electronic-connections}
\end{figure}

The tunneling current loop is made of coaxial cable and is completely decoupled from the rest of the STM system to avoid the negative influence of ground loops (see schematics in FIG.~\ref{fig:electronic-connections}). The bias voltage provided by the STM controller goes first through an amplifier (DATAFORTH SCM5B41-03)\cite{dataforth} with a gain factor of 1 and a band-width of 10 kHz, which galvanically separates the signal from the rest of the system. Thereafter, the signal is fed into a voltage divider that has the possibility to divide the signal by 1, 10 or 100. Finally the bias voltage is applied to the sample. The tunneling current is first received by a current amplifier (FEMTO DLPCA-200), which converts the current to a voltage signal with a gain factor of 1 nA/V. The band-width of this amplifier is of about 1.1 kHz\cite{femto}. The resulting signal is then sent to a second amplifier (DATAFORTH SCM5B41-03) and subsequently to the STM controller.

The electric ground of both galvanic amplifiers, on the STM side, are connected to a common reference point together with the shields of the isolated coaxial cables that go to the STM. On the other hand, on the STM controller side, the electric ground of the amplifiers shares a common reference point with the STM controller ground. 

\begin{figure}
\includegraphics[scale=0.44]{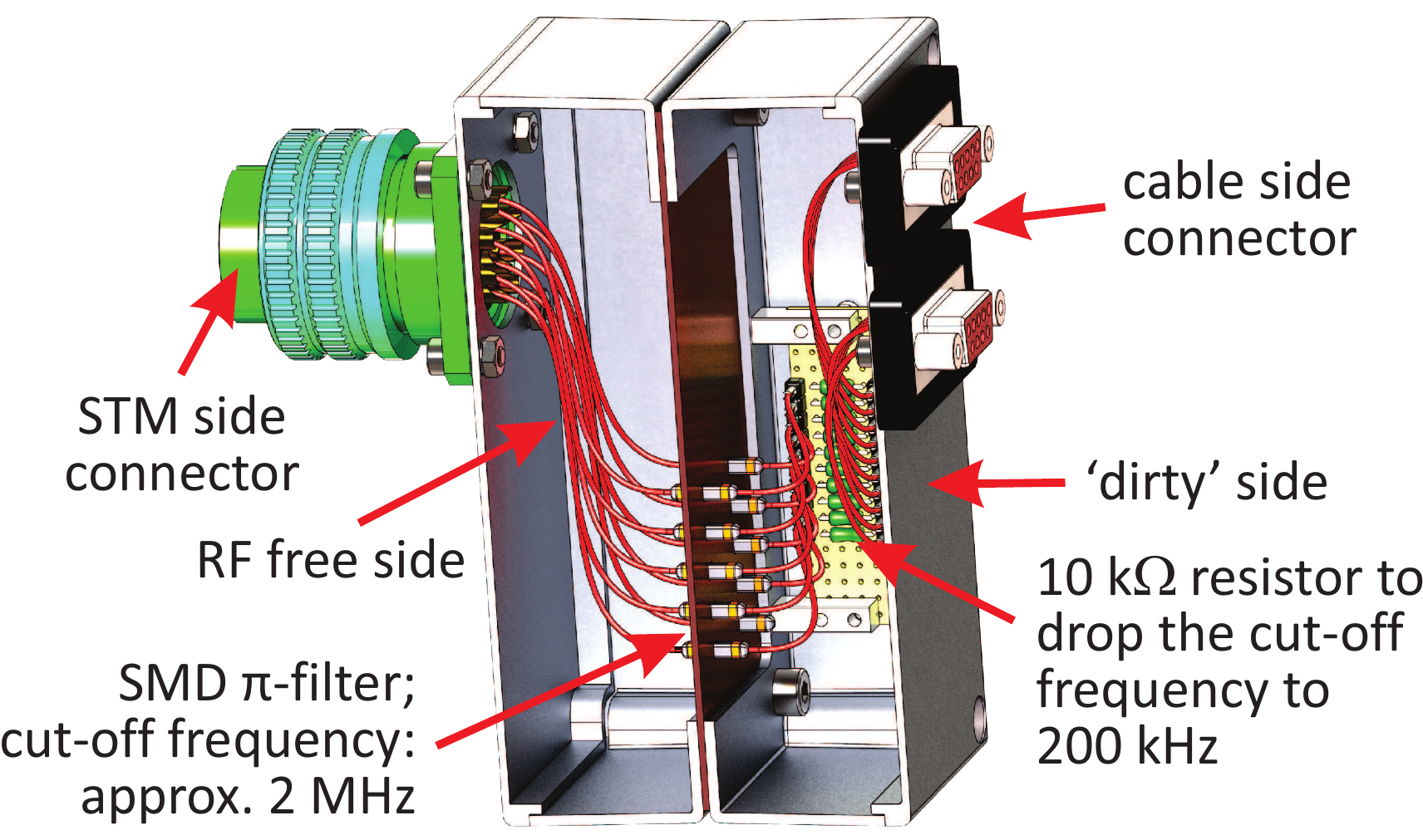}
\caption{RF-filter box with $\pi$-filters for each signal line to the $^3$He stage of the STM system. Dimensions: 60 mm x 63 mm x 111 mm. Reproduced from Reference 30.}
\label{fig:RF-filter}
\end{figure}

All the mentioned components, except the electronic control, are mounted in a radio frequency (RF) shielded metal-box. Furthermore, all the connecting cables to the STM as well as to the sensors and heaters in the $^3$He stage are protected with $\pi$-filter against RF-radiation\cite{assig201310, Morgenstern2017rsi}. The corresponding RF-filter boxes are mounted between the connecting wires and the UHV-feed-through flanges (FIG.~\ref{fig:RF-filter} shows an RF-filter box). Each of the boxes consists of two small chambers divided by a copper plate. The plates host the $\pi$-filters of the type TUSONIX 4701-001\cite{tusonix} with a cutting frequency of 2 MHz. In order to increase the filter effect, resistors of about 10 k$\Omega$ are installed in series. The cutting frequency can thus be set up to 200 kHz.

\subsection{UHV system}

The preparation and the exchange of tips/samples require proper UHV conditions outside the cryostat. Therefore, three UHV chambers are installed in the bottom part of the cryostat as is shown in FIG.~\ref{fig:overview}. These three chambers are inter-connected via two valves that let them operate independently. The tip/sample transfer is carried out by using magnetically coupled rotary-slide manipulators. The UHV is maintained by different pumps that will be detailed below for each specific chamber.

The STM chamber is directly mounted below the cryostat through a DN 350 CF-flange. The vacuum in this chamber is sustained by an ion getter pump with an ultimate pressure below 10$^{-11}$ mbar. The STM chamber contains in the central part the sample storage carousel, which additionally serves as bridge in the sample/tip transfer between the STM and the preparation and analysis chambers. The STM chamber has two manipulators: one wobble stick for the exchange of sample/tip between the STM and the carousel and a second manipulator, located in the bottom part, to open and close the window of the 1 K-shield, which protects the STM.

The preparation chamber offers the possibility to prepare the samples and tips for the SP-STM experiment. For this purpose, the chamber contains surface cleaning devices, in particular an argon-ion sputter gun and a high temperature heating stage for flashing tips and samples at temperatures approximately up to 2700~K. Furthermore, there is the option to install different evaporators for molecules and metals for the growth of thin films from the sub-monolayer coverage up to several layers on the sample or the tip. The latter constitutes a particularly important requirement for the preparation of samples and tips for spin resolved measurements. The chemical environment of the chamber is controlled via a quadrupole mass spectrometer with a 0-100 atomic mass unit range, while the vacuum is maintained by a combination of a dry scroll vacuum pump with ultimate pressure of 6.6x10$^{-2}$ mbar , a turbo molecular pump with a base pressure below 2x10$^{-10}$mbar, a titanium sublimation pump and an ion getter pump. An entry-lock chamber is connected to the preparation chamber for exchanging samples and tips between the external environment and the UHV-system. This small chamber has the option to be pumped and vented relatively fast through a second turbo molecular pump. The manipulation of samples and tips in the preparation and entry-lock chambers is carried out by four different manipulators: two of them are one-dimensional manipulators (one from the entry-lock chamber to the preparation chamber, a second from the preparation chamber to the STM chamber), a third one has an $xy$-table for positioning the sample/tip with more freedom in the space of the preparation chamber, and finally a wobble stick manipulator, which allows free three-dimensional movement in the preparation chamber.

The analysis chamber is designed for characterization of the sample surfaces by LEED (low energy electron diffraction) and elemental analysis through Auger spectroscopy. Further analysis and preparation tools such as evaporators, sputter gun, mass spectrometer, etc., can be installed if necessary. The vacuum in this chamber is maintained with a similar setup as in the preparation chamber. The chamber is furthermore equipped with one one-dimensional and another two-dimensional manipulator. Additionally, the chamber has a wobble stick manipulator for three-dimensional movements.  
            
\section{Demonstration of functionality}

\subsection{Tunneling current noise}

\begin{figure}
\includegraphics[scale=0.33]{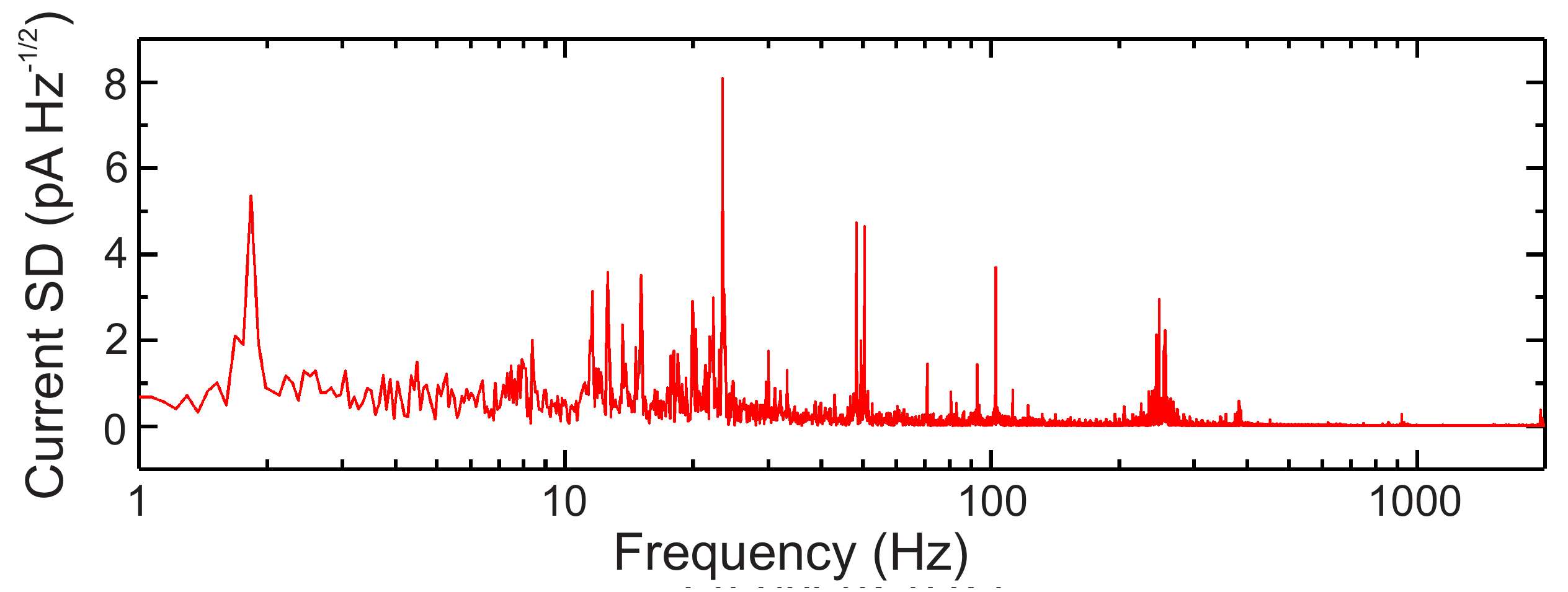}
\caption{Current spectral density noise numerically calculated from tunneling current data taken without feedback loop in 2048 x 128 measurement points. The sample was a W(110) single crystal and the tip was made of Nb. The current set point was 500 pA, the bias voltage -50 mV and the temperature 300 mK using the 1 K-pot.}
\label{fig:currentSD}
\end{figure} 
 
FIG.~\ref{fig:currentSD} shows the tunneling current spectral density noise from 1 Hz to 2000 Hz of a measurement without feedback loop at 300 mK employing the 1 K-pot. The spectrum depicts characteristic mechanical and electromagnetic noise of the STM system. In general for the low frequency regime, the spectral density background stays below 1 pA Hz$^{-1/2}$ and for high frequencies the spectral density background stays below 0.1 pA Hz$^{-1/2}$. The low frequency region shows a potential disturbing peak at about 23 Hz with the highest intensity. In order to overcome this potential disturbing peak, the measurement time per point (pixel) should be longer than 100 ms. Proper measurement settings let us obtain, for example, topographic data without the presence of disturbing frequencies as it is demonstrated in FIG.11.  

\subsection{Atomic resolution}

\begin{figure}
\includegraphics[scale=0.44]{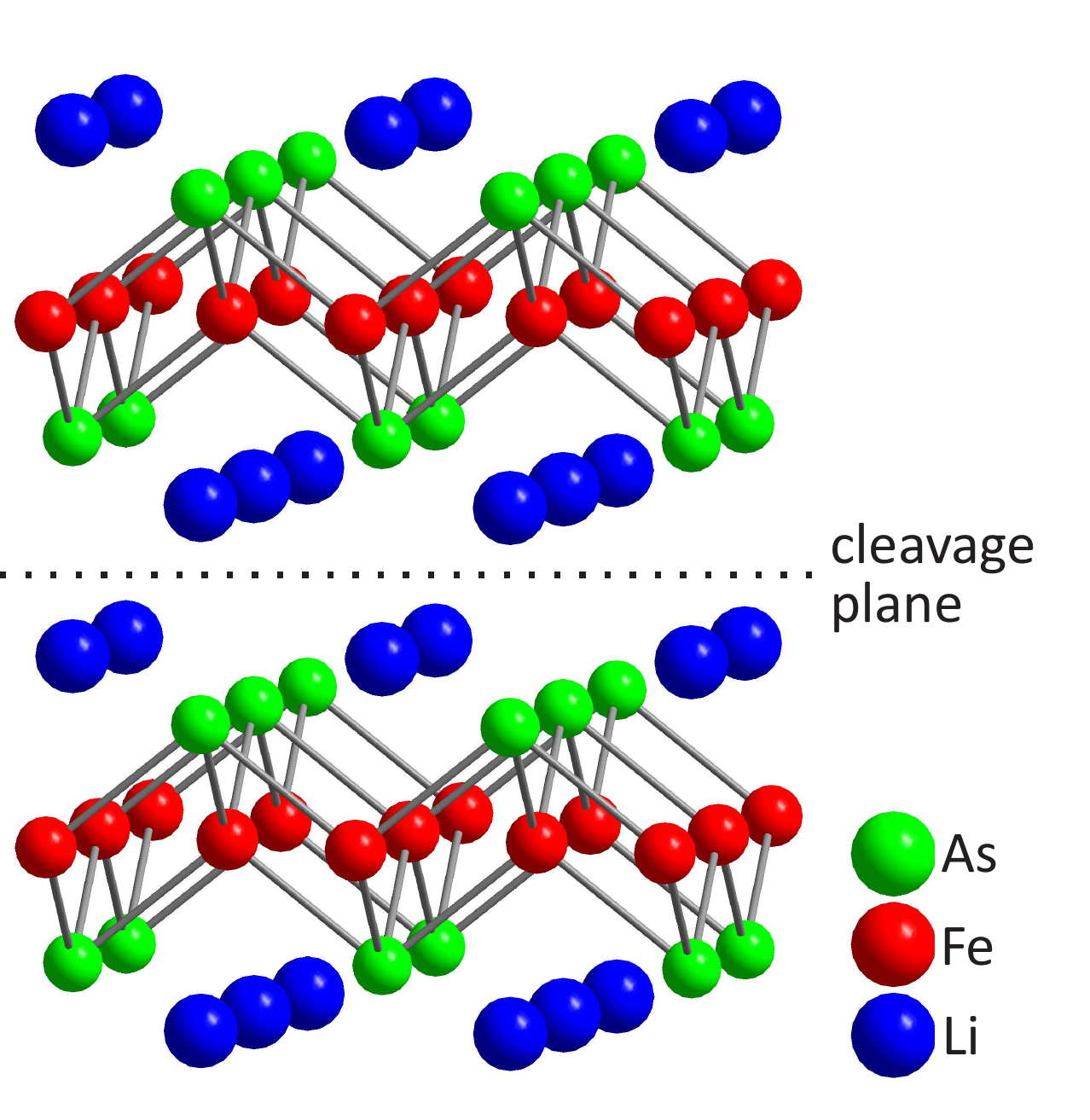}
\caption{Schematic of the LiFeAs structure. Fe in red, As in green and Li in blue. The cleaving plane is between the Li ions. Reproduced from Reference 30.}
\label{LiFeAs-1}
\end{figure}

In order to demonstrate the spatial resolution of the instrument, a sample of the type II superconducting material LiFeAs has been topographically investigated. The crystal was grown as described by Morozov et al.\cite{morozov2010single} and its structure is shown in FIG.~\ref{LiFeAs-1}. It is known that the material cleaves very well between two Li layers, resulting in high quality surfaces which can further show atomic resolution in STM measurements\cite{nag2016two, hanke2012probing,Schlegel2014,schlegel2016defect}.

\begin{figure}
\includegraphics[scale=0.44]{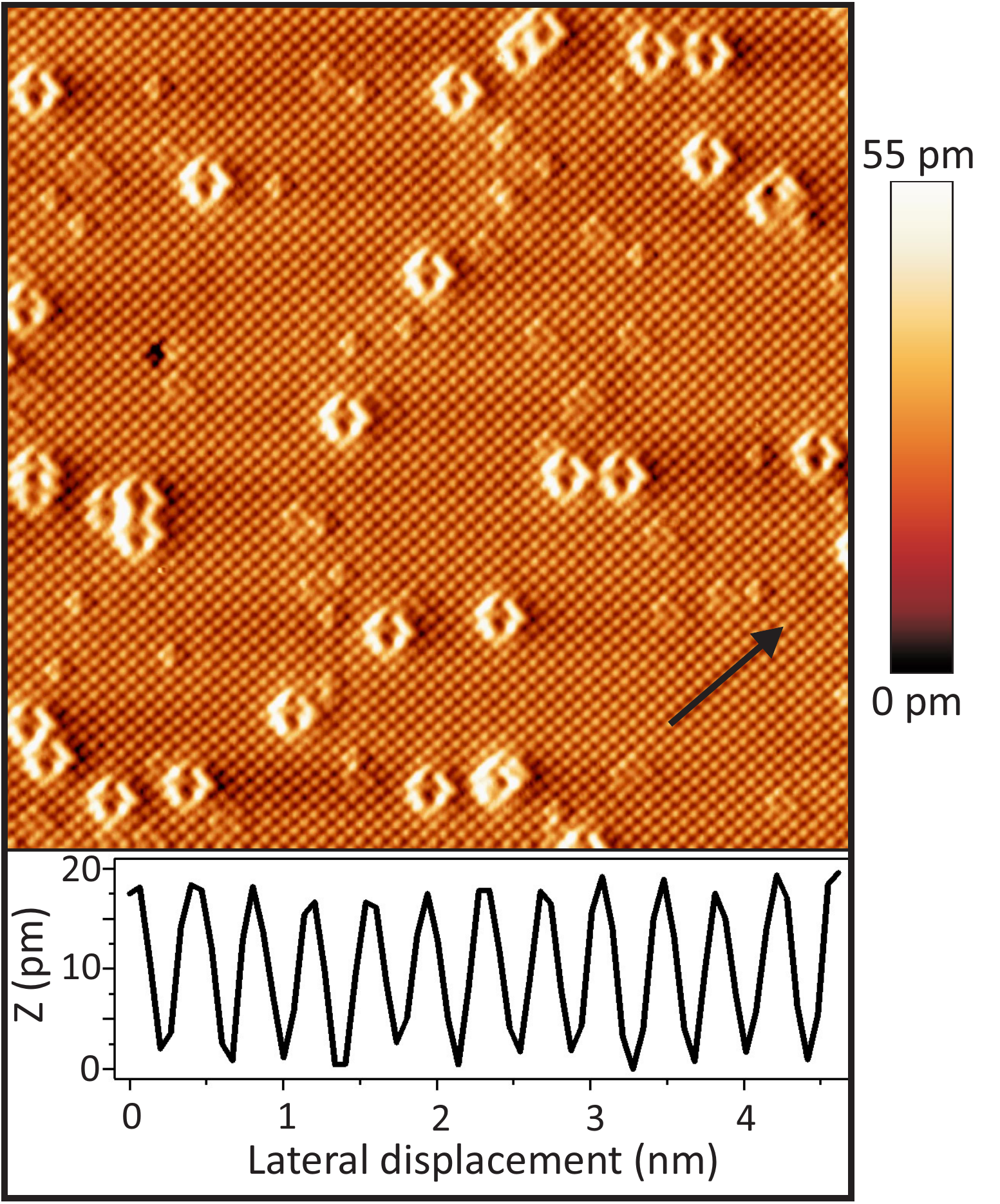}
\caption{Constant current topographic image of LiFeAs and line profile along black arrow. Area: 25.7 nm x 25.7 nm, $U = -50$~mV, $I = 2$~nA and $T= 5.8$~K. Reproduced from Reference 30.}
\label{LiFeAs}
\end{figure}   

FIG.~11 shows the topography of the sample measured at 5.8 K. The characteristic atomic corrugation of LiFeAs as well as several surface defects \cite{schlegel2016defect} are resolved with high resolution. The data prove atomic resolution and can be further used as reference for scanner calibration.
                          
\subsection{Energy resolution and temperature variability}

In order to demonstrate the high energy resolution and the temperature variability of the STM, spectroscopic measurements employing a Nb tip and a W(110) sample have been carried out. The Nb tips were mechanically shaped from polycrystalline Nb wires of 99.9\% purity, and the W(110) sample was prepared in UHV conditions through several cycles of annealing at 1500 K in oxygen atmosphere ($p_{O_2}=10^{-7}$~mbar) with a subsequent high temperature treatment at 2300 K\cite{bode2007preparation}. Once the Nb tip was able to reveal the well known step-like W(110) surface, single point spectra were measured.

\begin{figure}
\includegraphics[scale=0.5]{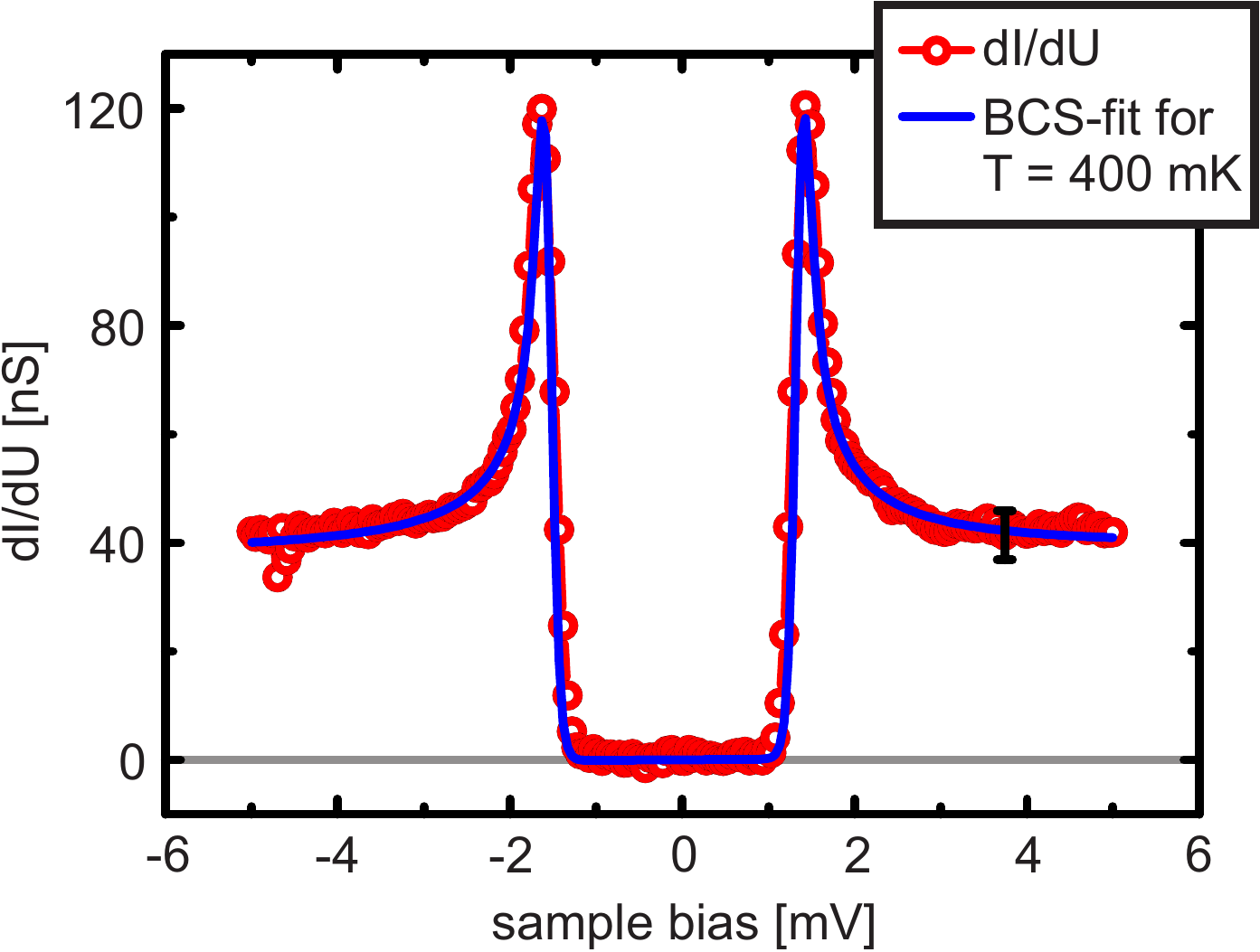}
\caption{STS at base temperature employing a Nb tip and a W(110) crystal (red circles correspond to experimental data and blue line to BCS-fit). Temperature at the STM $T_\mathrm{STM} = 378$~mK, temperature of the BCS fit  $T_\mathrm{fit} = 400$~mK, superconducting gap $\Delta = 1.45$~mV. Measurement conditions: bias voltage from $-5$~mV to $+5$~mV, $\Delta U = 50~\mu$V, $U_\mathrm{mod} = 20~\mu$$V_\mathrm{RMS}$, stabilized at $U_\mathrm{bias} = -5$~mV and $I = 200$pA. Reproduced from Reference 30.}
\label{energy-resolution}
\end{figure}

FIG.~\ref{energy-resolution} shows an averaged spectrum taken at 378 mK. The red circles correspond to the experimental data and the blue line to the BCS density of states\cite{PhysRev.106.162,Tinkham}  given by 

\begin{equation}
\frac{dI}{dU}\propto \int_{-\infty}^{\infty} \frac{|\epsilon|}{\sqrt{\epsilon^2-\Delta^2}} \cdot \frac{\frac{1}{k_B \cdot T} \cdot \exp(\frac{\epsilon + eU}{k_B \cdot T})}{(\exp(\frac{\epsilon + eU}{k_B \cdot T})+1)^2} d\epsilon,
\end{equation}

The density of states can be very well fitted, providing a value of $\Delta = 1.45$~mV for the superconducting gap of Nb, which is in very good agreement with literature\cite{wiebe2004300, pan1998vacuum, townsend1962investigation}. Additionally, the BCS fit yields an electronic temperature $T_\mathrm{fit} = 400$~mK, i.e. in very good agreement with the thermometer reading. Thus, the energy resolution is close to the theoretical limit of $\Delta E \approx 3.5k_BT\approx 120~\mu$eV.

\begin{figure}
\includegraphics[scale=0.5]{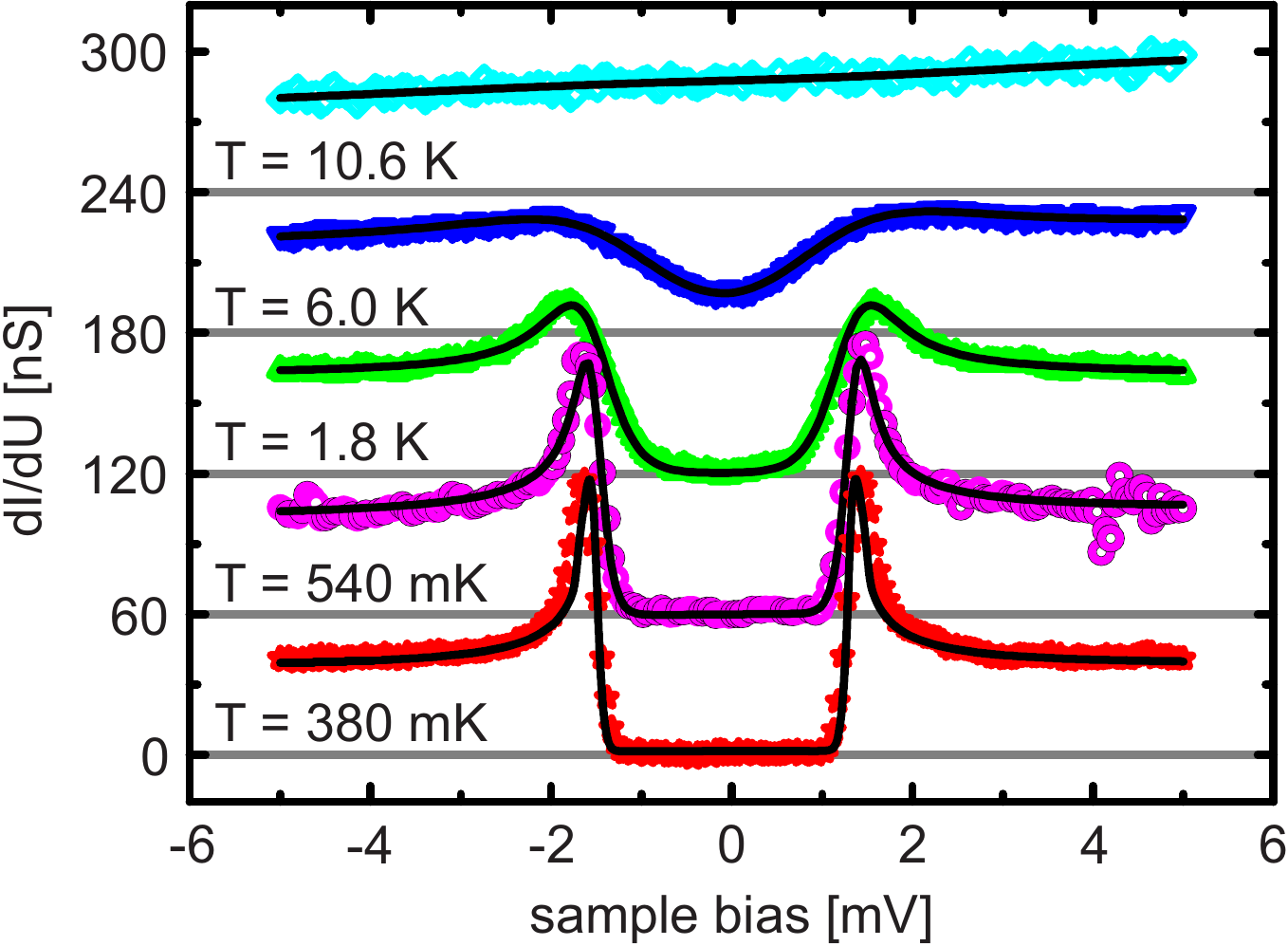}
\caption{Temperature dependent change of the superconducting gap of the Nb tip. Measurement conditions: bias voltage from -5 mV to +5 mV, $\Delta U = 50~\mu$V, $U_\mathrm{mod} = 50~\mu$$V_\mathrm{RMS}$, stabilized at $U_\mathrm{bias} = -5$~mV and $I = 200$~pA. Reproduced from Reference 30.}
\label{energy-resol-temp}
\end{figure}  

The capability to study $dI/dU$ spectroscopy at various temperatures was additionally tested. FIG.~\ref{energy-resol-temp} shows according temperature dependent spectroscopic measurements from 380 mK up to 10.6~K, where no superconducting gap is detected.

\subsection{Measurements in magnetic field}

\begin{figure}
\includegraphics[scale=0.5]{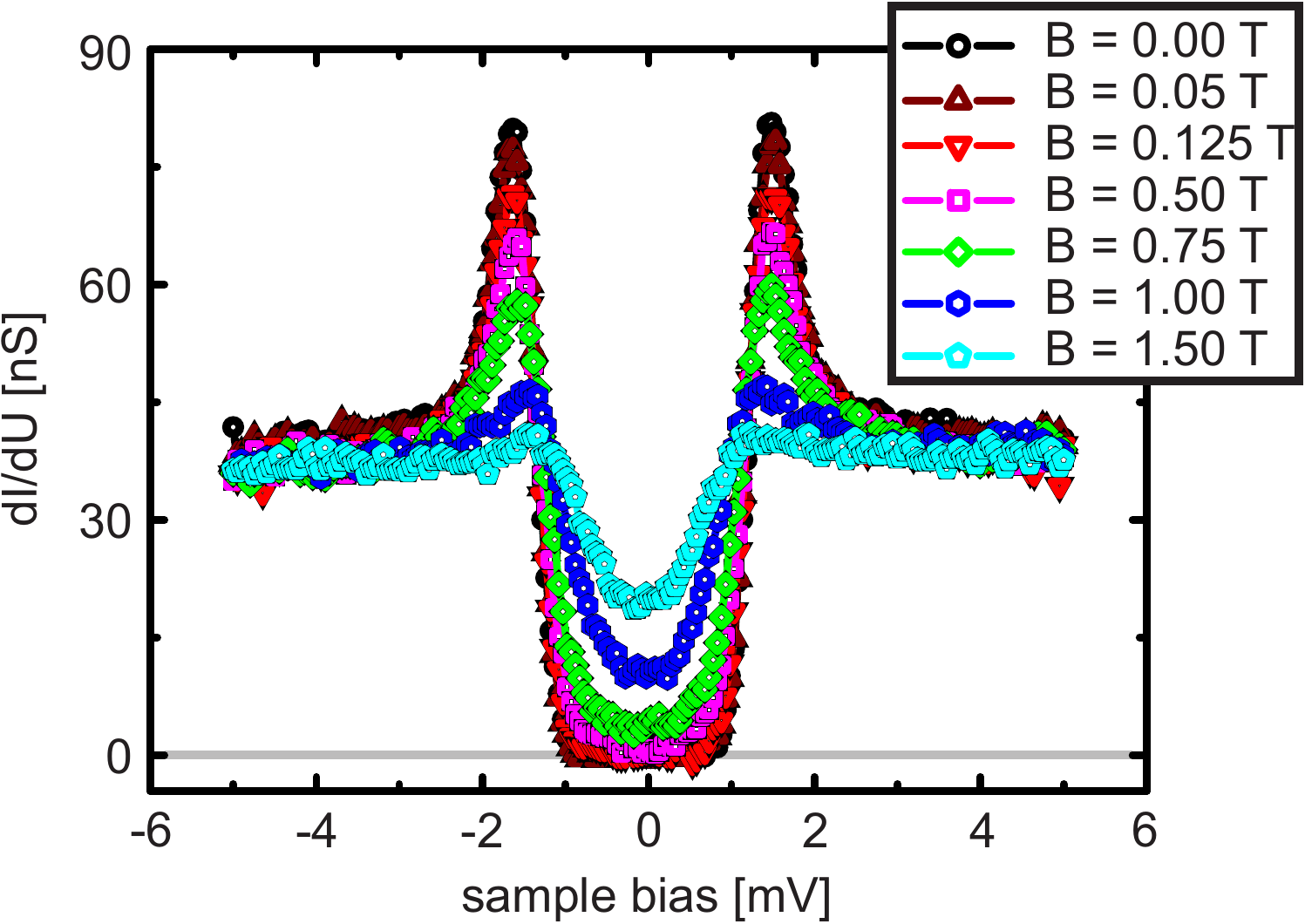}
\caption{Magnetic field dependent change of the superconducting gap of the Nb tip. Measurement conditions: bias voltage from -5 mV to +5 mV, $\Delta U = 50~\mu\mathrm{V}$, $U_\mathrm{mod} = 50~\mu$$V_\mathrm{RMS}$, stabilized at $U_{bias} = -5$~mV and $I = 200$~pA. Reproduced from Reference 30.}
\label{magnfield}
\end{figure}

As it has been described in section II-B, the system is equipped with a superconducting magnet which provides magnetic fields up to 9 T perpendicular to the sample surface. Accordingly, FIG.~\ref{magnfield} shows measurements of the differential conductance of the above discussed Nb-W tunneling junction for various magnetic fields. The reduction of the superconducting gap upon increasing the magnetic field from zero to 1.50~T is clearly visible.

\subsection{Spin-polarized measurements}

\begin{figure}
\includegraphics[scale=0.47]{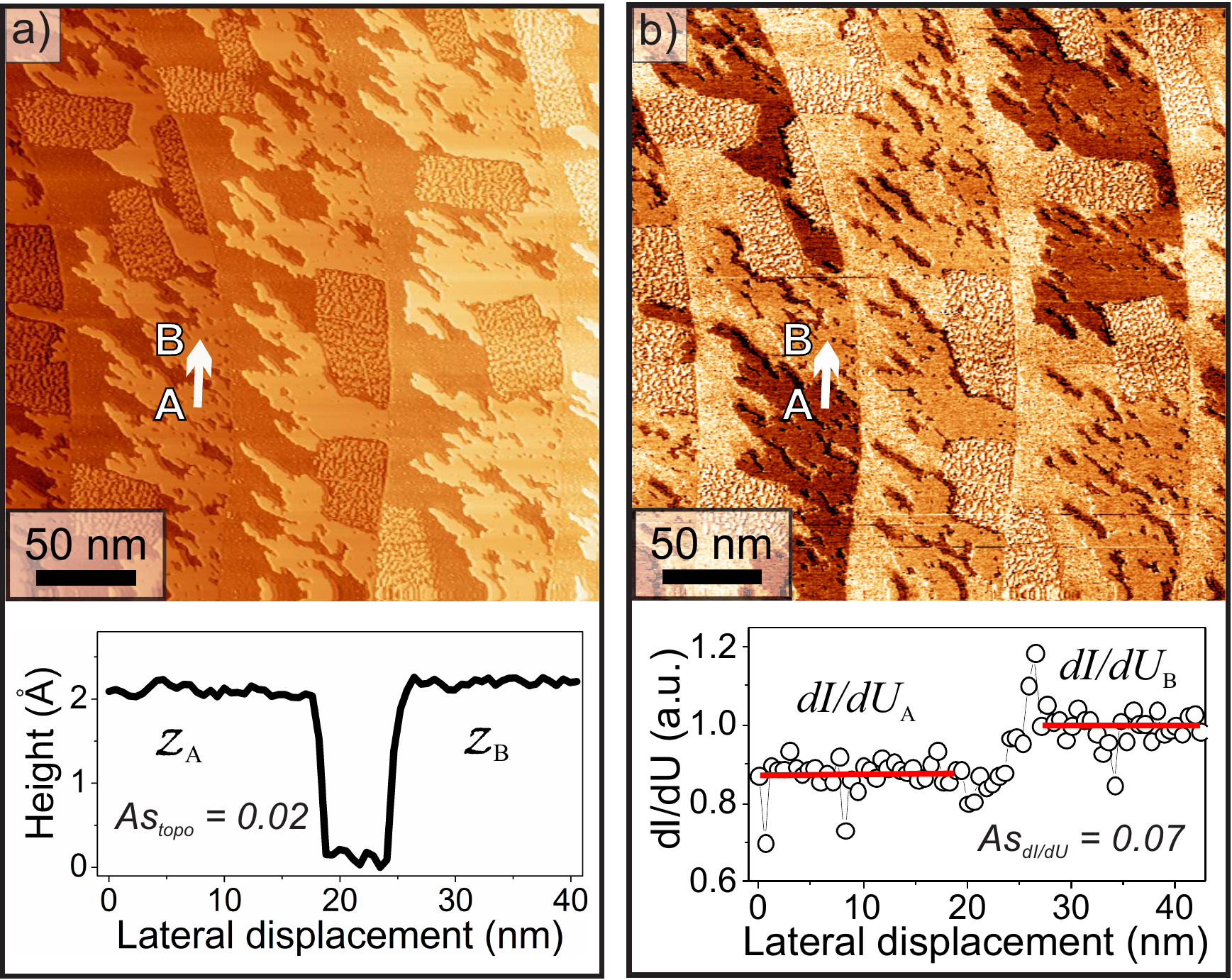}
\caption{Iron islands (1.6 ML Fe) on W(110) measured with an iron-coated tungsten tip. (a) Topographic image ($I = 300$~pA, $U = -300$~mV, $T = 38$~K), (bottom) line profile along white arrow (b) dI/dU-map @ $U = -300$~mV, (bottom) line profile along white arrow. Reproduced from Reference 31.}
\label{spinpol1}
\end{figure}

\begin{figure}
\includegraphics[scale=0.46]{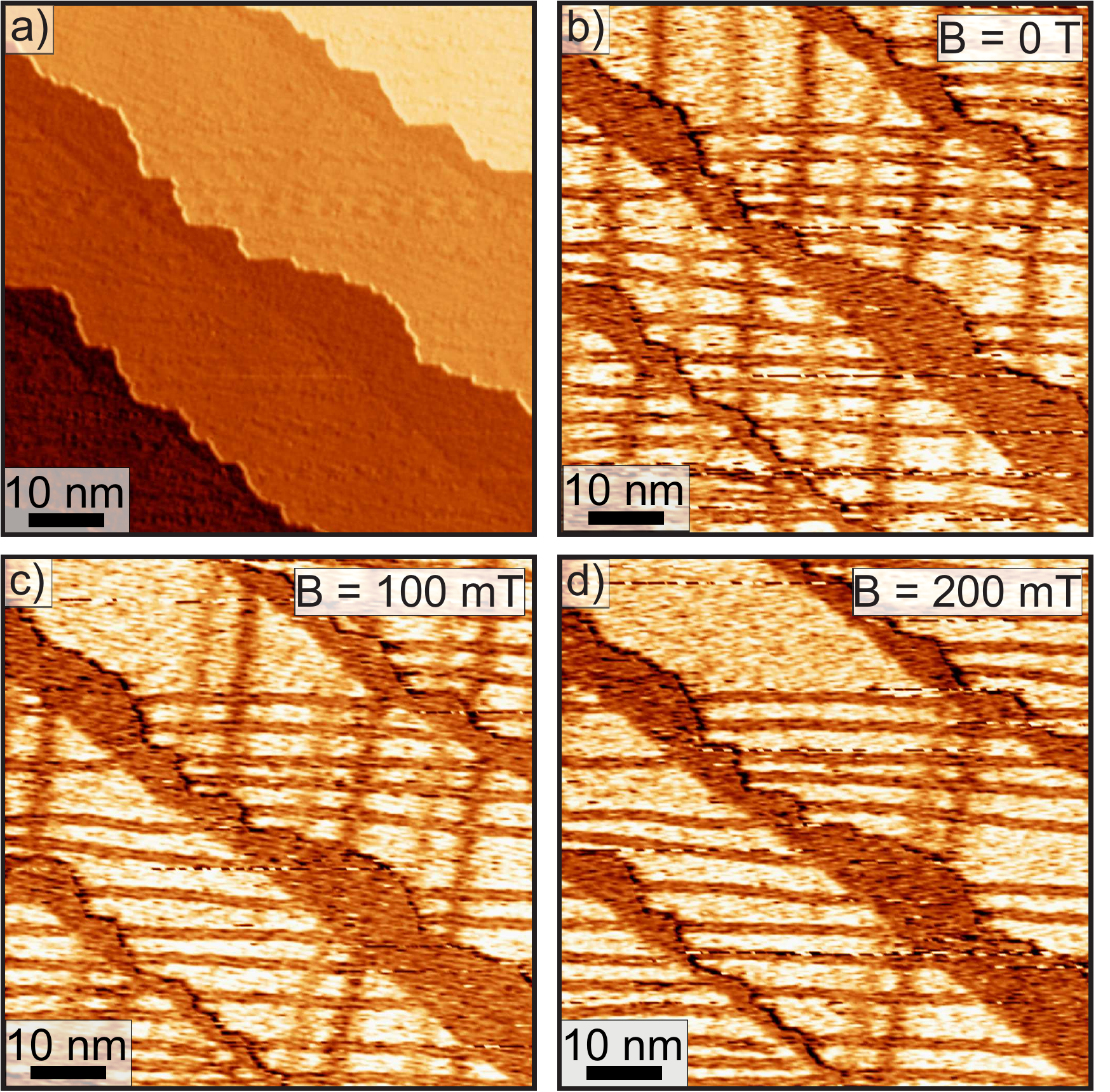}
\caption{Iron nanowires (1.5 ML Fe) on W(110) measured with a chromium bulk tip in magnetic field. (a) Topographic image ($I = 900$~pA, $U = +60$~mV, $T = 35$~K), (b) dI/dU-map @ $U = +60$~mV, $B = 0$~T, (c) dI/dU-map @ $U = +60$~mV, $B = 100$~mT, (d) dI/dU-map @ $U = +60$~mV, $B = 200$~mT. Reproduced from Reference 32.}
\label{spinpol2}
\end{figure}

One of the main characteristics of the STM system is the capability to perform spin-resolved measurements. SP-STM relies on the tunneling magnetoresistance effect, which means that the relative orientation of the magnetic moments of tip and sample determines the value of the tunneling current\cite{wiesendanger2009spin, bode2003spin}. To demonstrate the capability of spin resolution, the magnetic structure of iron nanostructures has been studied using magnetic tunneling tips. The samples were prepared under UHV conditions by evaporating between one and two atomic layers of iron on a W(110) single crystal. Hereby, different growth factors such as annealing temperature, annealing time, substrate defects, etc. were determinant for the obtained iron nanostructures\cite{Kirsten2004, Salazar2016, Scheffler2015}. Two different samples were prepared: FIG.~\ref{spinpol1} shows the results for iron islands of two atomic layers height, while FIG.~\ref{spinpol2} presents results for iron nanowires of two atomic layers height. 

For the two-monolayer islands of the type shown in FIG.~\ref{spinpol1} it is well known that they exhibit an out of the plane spin polarization\cite{bode2003spin, wiesendanger2009spin}. For studying these islands,  an iron coated tungsten tip was used that was prepared by evaporating about 20 atomic layers of iron on a clean tungsten tip. Prior to that, the tungsten tip was prepared via chemical etching and subsequent high temperature treatments at 2000~K for 5 seconds using an electron beam heater.  A constant current topographic image of the sample is shown in FIG.~\ref{spinpol1} (a). In this case, islands of different sizes are located on the steps of the underlying tungsten crystal. The bottom of FIG.~\ref{spinpol1} (a) depicts a line profile taken along the white arrow shown in the topography. This line profile corresponds to the apparent height in two different islands labeled A and B (see Figure (a)). The average topographic height of island A ($z_A$) is similar to that of island B ($z_B$). The height asymmetry defined as $\text{As}_\mathrm{topo} = (z_B-z_A)/(z_B+z_A)$  is useful to evaluate the difference between two topographic signals of the two islands\cite{wiesendanger2009spin}. Here, we find $\text{As}_\mathrm{topo} = 0.02$, which indicates a tiny difference of both islands. FIG.~\ref{spinpol1}(b) shows a $dI/dU$-map of the same sample taken at $U=-300$~mV. The islands in this case show evidently different differential conductance, because some of them appear bright and others dark. This significant contrast very clearly demonstrates a spin-polarized tunneling current and thus sensitivity to the magnetic structure of the sample\cite{kubetzka2001magnetism, wiesendanger2009spin}. This is confirmed by the asymmetry of the differential conductance $\text{As}_{dI/dU}=(dI/dU_B-dI/dU_A)/(dI/dU_B+dI/dU_A)=0.07$.

FIG.~\ref{spinpol2}(a) shows the topography of iron nanowires grown via the step flow mechanism\cite{bode2003spin}. The sample was measured with a mechanically sharpened chromium bulk tip, which was made using a chromium single crystal of high purity (approximately 99\%) and additionally evaporating 10 to 100 monolayers of fresh/clean Chromium via MBE. In the $dI/dU$-maps shown in FIGs.~\ref{spinpol2}(b,c,d), one can distinguish the presence of two types of lines: i) a group of quasi-horizontal lines with respect to the image frame and ii) a second group of quasi-vertical lines with respect to the image frame. The horizontal lines correspond to dislocation lines produced by the strain caused by the different lattice parameters between iron and tungsten\cite{Kirsten2004}. The vertical lines with alternate intensities (bright and dark) are domain walls of the magnetic structure along the double layer nanowires as it has been previously reported\cite{bode2003spin}. The alternate intensity of the domain walls indicate a rotational sense of the magnetic domains. It is worth to mention that here the tip possess an in-plane magnetic sensitivity, therefore the magnetic contrast of the out-of-plane magnetized domains is not visible. By applying an external magnetic field as is shown in FIGs.~\ref{spinpol2}(b,c,d), the magnetic domains within the nanowires become alternatively larger and smaller upon increasing the magnetic field. Concomitantly, the dark and bright vertical lines, i.e. the domain walls, are pairwise moved against each other.

\section{Summary}

The construction and the operation of a robust STM system is reported. The system's special design combines ultra-low temperature, high magnetic field and ultra-high vacuum techniques with the sensitive STM technique. Considering its outstanding resolution and the UHV facilities, the STM is able to perform spin-resolved STM/STS measurements. The functionality of the system has been probed by measuring different well known samples such as: i) a cleavable LiFeAs single crystal with a tungsten tip, ii) a tungsten single crystal with a niobium tip and iii) two atomic layer iron nanostructures on W(110) with iron-coated tugnsten tips as well as chromiun bulk tips.

\begin{acknowledgments}
The authors thank the mechanical and electrical workshop of the IFW Dresden for support. Further, this project has been supported by the Deutsche Forschungsgemeinschaft through the Priority Programme SPP1458 (Grant HE3439/11), the Research Unit 1154 (Grant HE3439/10), the Collaborative Research Center SFB 1143 (Project C05), and through GRK 1621. Furthermore, this project has received funding from the European Research Council (ERC) under the European Unions' Horizon 2020 research and innovation programme (grant agreement No 647276 -- MARS -- ERC-2014-CoG).
\end{acknowledgments}



\begin{thebibliography}{56}%
\makeatletter
\providecommand \@ifxundefined [1]{%
 \@ifx{#1\undefined}
}%
\providecommand \@ifnum [1]{%
 \ifnum #1\expandafter \@firstoftwo
 \else \expandafter \@secondoftwo
 \fi
}%
\providecommand \@ifx [1]{%
 \ifx #1\expandafter \@firstoftwo
 \else \expandafter \@secondoftwo
 \fi
}%
\providecommand \natexlab [1]{#1}%
\providecommand \enquote  [1]{``#1''}%
\providecommand \bibnamefont  [1]{#1}%
\providecommand \bibfnamefont [1]{#1}%
\providecommand \citenamefont [1]{#1}%
\providecommand \href@noop [0]{\@secondoftwo}%
\providecommand \href [0]{\begingroup \@sanitize@url \@href}%
\providecommand \@href[1]{\@@startlink{#1}\@@href}%
\providecommand \@@href[1]{\endgroup#1\@@endlink}%
\providecommand \@sanitize@url [0]{\catcode `\\12\catcode `\$12\catcode
  `\&12\catcode `\#12\catcode `\^12\catcode `\_12\catcode `\%12\relax}%
\providecommand \@@startlink[1]{}%
\providecommand \@@endlink[0]{}%
\providecommand \url  [0]{\begingroup\@sanitize@url \@url }%
\providecommand \@url [1]{\endgroup\@href {#1}{\urlprefix }}%
\providecommand \urlprefix  [0]{URL }%
\providecommand \Eprint [0]{\href }%
\providecommand \doibase [0]{http://dx.doi.org/}%
\providecommand \selectlanguage [0]{\@gobble}%
\providecommand \bibinfo  [0]{\@secondoftwo}%
\providecommand \bibfield  [0]{\@secondoftwo}%
\providecommand \translation [1]{[#1]}%
\providecommand \BibitemOpen [0]{}%
\providecommand \bibitemStop [0]{}%
\providecommand \bibitemNoStop [0]{.\EOS\space}%
\providecommand \EOS [0]{\spacefactor3000\relax}%
\providecommand \BibitemShut  [1]{\csname bibitem#1\endcsname}%
\let\auto@bib@innerbib\@empty
\bibitem [{\citenamefont {Crommie}, \citenamefont {Lutz},\ and\ \citenamefont
  {Eigler}(1993)}]{crommie1993confinement}%
  \BibitemOpen
  \bibfield  {author} {\bibinfo {author} {\bibfnamefont {M.~F.}\ \bibnamefont
  {Crommie}}, \bibinfo {author} {\bibfnamefont {C.~P.}\ \bibnamefont {Lutz}}, \
  and\ \bibinfo {author} {\bibfnamefont {D.~M.}\ \bibnamefont {Eigler}},\
  }\href@noop {} {\bibfield  {journal} {\bibinfo  {journal} {Science}\ }\textbf
  {\bibinfo {volume} {262}},\ \bibinfo {pages} {218} (\bibinfo {year}
  {1993})}\BibitemShut {NoStop}%
\bibitem [{\citenamefont {Yazdani}\ \emph {et~al.}(1997)\citenamefont
  {Yazdani}, \citenamefont {Jones}, \citenamefont {Lutz}, \citenamefont
  {Crommie},\ and\ \citenamefont {Eigler}}]{Yazdani1997}%
  \BibitemOpen
  \bibfield  {author} {\bibinfo {author} {\bibfnamefont {A.}~\bibnamefont
  {Yazdani}}, \bibinfo {author} {\bibfnamefont {B.~A.}\ \bibnamefont {Jones}},
  \bibinfo {author} {\bibfnamefont {C.~P.}\ \bibnamefont {Lutz}}, \bibinfo
  {author} {\bibfnamefont {M.~F.}\ \bibnamefont {Crommie}}, \ and\ \bibinfo
  {author} {\bibfnamefont {D.~M.}\ \bibnamefont {Eigler}},\ }\href {\doibase
  10.1126/science.275.5307.1767} {\bibfield  {journal} {\bibinfo  {journal}
  {Science}\ }\textbf {\bibinfo {volume} {275}},\ \bibinfo {pages} {1767}
  (\bibinfo {year} {1997})}\BibitemShut {NoStop}%
\bibitem [{\citenamefont {Hoffman}\ \emph {et~al.}(2002)\citenamefont
  {Hoffman}, \citenamefont {McElroy}, \citenamefont {Lee}, \citenamefont
  {Lang}, \citenamefont {Eisaki}, \citenamefont {Uchida},\ and\ \citenamefont
  {Davis}}]{Hoffman2002}%
  \BibitemOpen
  \bibfield  {author} {\bibinfo {author} {\bibfnamefont {J.~E.}\ \bibnamefont
  {Hoffman}}, \bibinfo {author} {\bibfnamefont {K.}~\bibnamefont {McElroy}},
  \bibinfo {author} {\bibfnamefont {D.-H.}\ \bibnamefont {Lee}}, \bibinfo
  {author} {\bibfnamefont {K.~M.}\ \bibnamefont {Lang}}, \bibinfo {author}
  {\bibfnamefont {H.}~\bibnamefont {Eisaki}}, \bibinfo {author} {\bibfnamefont
  {S.}~\bibnamefont {Uchida}}, \ and\ \bibinfo {author} {\bibfnamefont {J.~C.}\
  \bibnamefont {Davis}},\ }\href
  {http://www.sciencemag.org/content/297/5584/1148.abstract} {\bibfield
  {journal} {\bibinfo  {journal} {Science}\ }\textbf {\bibinfo {volume}
  {297}},\ \bibinfo {pages} {1148} (\bibinfo {year} {2002})}\BibitemShut
  {NoStop}%
\bibitem [{\citenamefont {Yee}\ \emph {et~al.}(2015)\citenamefont {Yee},
  \citenamefont {Zhu}, \citenamefont {Soumyanarayanan}, \citenamefont {He},
  \citenamefont {Song}, \citenamefont {Pomjakushina}, \citenamefont {Salman},
  \citenamefont {Kanigel}, \citenamefont {Segawa}, \citenamefont {Ando},\ and\
  \citenamefont {Hoffman}}]{yee2015spin}%
  \BibitemOpen
  \bibfield  {author} {\bibinfo {author} {\bibfnamefont {M.~M.}\ \bibnamefont
  {Yee}}, \bibinfo {author} {\bibfnamefont {Z.-H.}\ \bibnamefont {Zhu}},
  \bibinfo {author} {\bibfnamefont {A.}~\bibnamefont {Soumyanarayanan}},
  \bibinfo {author} {\bibfnamefont {Y.}~\bibnamefont {He}}, \bibinfo {author}
  {\bibfnamefont {C.-L.}\ \bibnamefont {Song}}, \bibinfo {author}
  {\bibfnamefont {E.}~\bibnamefont {Pomjakushina}}, \bibinfo {author}
  {\bibfnamefont {Z.}~\bibnamefont {Salman}}, \bibinfo {author} {\bibfnamefont
  {A.}~\bibnamefont {Kanigel}}, \bibinfo {author} {\bibfnamefont
  {K.}~\bibnamefont {Segawa}}, \bibinfo {author} {\bibfnamefont
  {Y.}~\bibnamefont {Ando}}, \ and\ \bibinfo {author} {\bibfnamefont
  {J.}~\bibnamefont {Hoffman}},\ }\href@noop {} {\bibfield  {journal} {\bibinfo
   {journal} {Physical Review B}\ }\textbf {\bibinfo {volume} {91}},\ \bibinfo
  {pages} {161306} (\bibinfo {year} {2015})}\BibitemShut {NoStop}%
\bibitem [{\citenamefont {Fu}\ \emph {et~al.}(2014)\citenamefont {Fu},
  \citenamefont {Kawamura}, \citenamefont {Igarashi}, \citenamefont {Takagi},
  \citenamefont {Hanaguri},\ and\ \citenamefont {Sasagawa}}]{fu2014imaging}%
  \BibitemOpen
  \bibfield  {author} {\bibinfo {author} {\bibfnamefont {Y.-S.}\ \bibnamefont
  {Fu}}, \bibinfo {author} {\bibfnamefont {M.}~\bibnamefont {Kawamura}},
  \bibinfo {author} {\bibfnamefont {K.}~\bibnamefont {Igarashi}}, \bibinfo
  {author} {\bibfnamefont {H.}~\bibnamefont {Takagi}}, \bibinfo {author}
  {\bibfnamefont {T.}~\bibnamefont {Hanaguri}}, \ and\ \bibinfo {author}
  {\bibfnamefont {T.}~\bibnamefont {Sasagawa}},\ }\href@noop {} {\bibfield
  {journal} {\bibinfo  {journal} {Nature Physics}\ }\textbf {\bibinfo {volume}
  {10}},\ \bibinfo {pages} {815} (\bibinfo {year} {2014})}\BibitemShut
  {NoStop}%
\bibitem [{\citenamefont {Kohsaka}\ \emph {et~al.}(2012)\citenamefont
  {Kohsaka}, \citenamefont {Hanaguri}, \citenamefont {Azuma}, \citenamefont
  {Takano}, \citenamefont {Davis},\ and\ \citenamefont
  {Takagi}}]{kohsaka2012visualization}%
  \BibitemOpen
  \bibfield  {author} {\bibinfo {author} {\bibfnamefont {Y.}~\bibnamefont
  {Kohsaka}}, \bibinfo {author} {\bibfnamefont {T.}~\bibnamefont {Hanaguri}},
  \bibinfo {author} {\bibfnamefont {M.}~\bibnamefont {Azuma}}, \bibinfo
  {author} {\bibfnamefont {M.}~\bibnamefont {Takano}}, \bibinfo {author}
  {\bibfnamefont {J.}~\bibnamefont {Davis}}, \ and\ \bibinfo {author}
  {\bibfnamefont {H.}~\bibnamefont {Takagi}},\ }\href@noop {} {\bibfield
  {journal} {\bibinfo  {journal} {Nature physics}\ }\textbf {\bibinfo {volume}
  {8}},\ \bibinfo {pages} {534} (\bibinfo {year} {2012})}\BibitemShut {NoStop}%
\bibitem [{\citenamefont {H{\"a}nke}\ \emph {et~al.}(2012)\citenamefont
  {H{\"a}nke}, \citenamefont {Sykora}, \citenamefont {Schlegel}, \citenamefont
  {Baumann}, \citenamefont {Harnagea}, \citenamefont {Wurmehl}, \citenamefont
  {Daghofer}, \citenamefont {B{\"u}chner}, \citenamefont {van~den Brink},\ and\
  \citenamefont {Hess}}]{hanke2012probing}%
  \BibitemOpen
  \bibfield  {author} {\bibinfo {author} {\bibfnamefont {T.}~\bibnamefont
  {H{\"a}nke}}, \bibinfo {author} {\bibfnamefont {S.}~\bibnamefont {Sykora}},
  \bibinfo {author} {\bibfnamefont {R.}~\bibnamefont {Schlegel}}, \bibinfo
  {author} {\bibfnamefont {D.}~\bibnamefont {Baumann}}, \bibinfo {author}
  {\bibfnamefont {L.}~\bibnamefont {Harnagea}}, \bibinfo {author}
  {\bibfnamefont {S.}~\bibnamefont {Wurmehl}}, \bibinfo {author} {\bibfnamefont
  {M.}~\bibnamefont {Daghofer}}, \bibinfo {author} {\bibfnamefont
  {B.}~\bibnamefont {B{\"u}chner}}, \bibinfo {author} {\bibfnamefont
  {J.}~\bibnamefont {van~den Brink}}, \ and\ \bibinfo {author} {\bibfnamefont
  {C.}~\bibnamefont {Hess}},\ }\href@noop {} {\bibfield  {journal} {\bibinfo
  {journal} {Physical review letters}\ }\textbf {\bibinfo {volume} {108}},\
  \bibinfo {pages} {127001} (\bibinfo {year} {2012})}\BibitemShut {NoStop}%
\bibitem [{\citenamefont {Bode}(2003)}]{bode2003spin}%
  \BibitemOpen
  \bibfield  {author} {\bibinfo {author} {\bibfnamefont {M.}~\bibnamefont
  {Bode}},\ }\href@noop {} {\bibfield  {journal} {\bibinfo  {journal} {Reports
  on Progress in Physics}\ }\textbf {\bibinfo {volume} {66}},\ \bibinfo {pages}
  {523} (\bibinfo {year} {2003})}\BibitemShut {NoStop}%
\bibitem [{\citenamefont {Wiesendanger}(2009)}]{wiesendanger2009spin}%
  \BibitemOpen
  \bibfield  {author} {\bibinfo {author} {\bibfnamefont {R.}~\bibnamefont
  {Wiesendanger}},\ }\href@noop {} {\bibfield  {journal} {\bibinfo  {journal}
  {Reviews of Modern Physics}\ }\textbf {\bibinfo {volume} {81}},\ \bibinfo
  {pages} {1495} (\bibinfo {year} {2009})}\BibitemShut {NoStop}%
\bibitem [{\citenamefont {Kamlapure}\ \emph {et~al.}(2013)\citenamefont
  {Kamlapure}, \citenamefont {Saraswat}, \citenamefont {Ganguli}, \citenamefont
  {Bagwe}, \citenamefont {Raychaudhuri},\ and\ \citenamefont
  {Pai}}]{kamlapure2013350}%
  \BibitemOpen
  \bibfield  {author} {\bibinfo {author} {\bibfnamefont {A.}~\bibnamefont
  {Kamlapure}}, \bibinfo {author} {\bibfnamefont {G.}~\bibnamefont {Saraswat}},
  \bibinfo {author} {\bibfnamefont {S.~C.}\ \bibnamefont {Ganguli}}, \bibinfo
  {author} {\bibfnamefont {V.}~\bibnamefont {Bagwe}}, \bibinfo {author}
  {\bibfnamefont {P.}~\bibnamefont {Raychaudhuri}}, \ and\ \bibinfo {author}
  {\bibfnamefont {S.~P.}\ \bibnamefont {Pai}},\ }\href@noop {} {\bibfield
  {journal} {\bibinfo  {journal} {Review of Scientific Instruments}\ }\textbf
  {\bibinfo {volume} {84}},\ \bibinfo {pages} {123905} (\bibinfo {year}
  {2013})}\BibitemShut {NoStop}%
\bibitem [{\citenamefont {Wiebe}\ \emph {et~al.}(2004)\citenamefont {Wiebe},
  \citenamefont {Wachowiak}, \citenamefont {Meier}, \citenamefont {Haude},
  \citenamefont {Foster}, \citenamefont {Morgenstern},\ and\ \citenamefont
  {Wiesendanger}}]{wiebe2004300}%
  \BibitemOpen
  \bibfield  {author} {\bibinfo {author} {\bibfnamefont {J.}~\bibnamefont
  {Wiebe}}, \bibinfo {author} {\bibfnamefont {A.}~\bibnamefont {Wachowiak}},
  \bibinfo {author} {\bibfnamefont {F.}~\bibnamefont {Meier}}, \bibinfo
  {author} {\bibfnamefont {D.}~\bibnamefont {Haude}}, \bibinfo {author}
  {\bibfnamefont {T.}~\bibnamefont {Foster}}, \bibinfo {author} {\bibfnamefont
  {M.}~\bibnamefont {Morgenstern}}, \ and\ \bibinfo {author} {\bibfnamefont
  {R.}~\bibnamefont {Wiesendanger}},\ }\href@noop {} {\bibfield  {journal}
  {\bibinfo  {journal} {Review of scientific instruments}\ }\textbf {\bibinfo
  {volume} {75}},\ \bibinfo {pages} {4871} (\bibinfo {year}
  {2004})}\BibitemShut {NoStop}%
\bibitem [{\citenamefont {Assig}\ \emph {et~al.}(2013)\citenamefont {Assig},
  \citenamefont {Etzkorn}, \citenamefont {Enders}, \citenamefont {Stiepany},
  \citenamefont {Ast},\ and\ \citenamefont {Kern}}]{assig201310}%
  \BibitemOpen
  \bibfield  {author} {\bibinfo {author} {\bibfnamefont {M.}~\bibnamefont
  {Assig}}, \bibinfo {author} {\bibfnamefont {M.}~\bibnamefont {Etzkorn}},
  \bibinfo {author} {\bibfnamefont {A.}~\bibnamefont {Enders}}, \bibinfo
  {author} {\bibfnamefont {W.}~\bibnamefont {Stiepany}}, \bibinfo {author}
  {\bibfnamefont {C.~R.}\ \bibnamefont {Ast}}, \ and\ \bibinfo {author}
  {\bibfnamefont {K.}~\bibnamefont {Kern}},\ }\href@noop {} {\bibfield
  {journal} {\bibinfo  {journal} {Review of Scientific Instruments}\ }\textbf
  {\bibinfo {volume} {84}},\ \bibinfo {pages} {033903} (\bibinfo {year}
  {2013})}\BibitemShut {NoStop}%
\bibitem [{\citenamefont {Roychowdhury}\ \emph {et~al.}(2014)\citenamefont
  {Roychowdhury}, \citenamefont {Gubrud}, \citenamefont {Dana}, \citenamefont
  {Anderson}, \citenamefont {Lobb}, \citenamefont {Wellstood},\ and\
  \citenamefont {Dreyer}}]{roychowdhury201430}%
  \BibitemOpen
  \bibfield  {author} {\bibinfo {author} {\bibfnamefont {A.}~\bibnamefont
  {Roychowdhury}}, \bibinfo {author} {\bibfnamefont {M.}~\bibnamefont
  {Gubrud}}, \bibinfo {author} {\bibfnamefont {R.}~\bibnamefont {Dana}},
  \bibinfo {author} {\bibfnamefont {J.}~\bibnamefont {Anderson}}, \bibinfo
  {author} {\bibfnamefont {C.}~\bibnamefont {Lobb}}, \bibinfo {author}
  {\bibfnamefont {F.}~\bibnamefont {Wellstood}}, \ and\ \bibinfo {author}
  {\bibfnamefont {M.}~\bibnamefont {Dreyer}},\ }\href@noop {} {\bibfield
  {journal} {\bibinfo  {journal} {Review of Scientific Instruments}\ }\textbf
  {\bibinfo {volume} {85}},\ \bibinfo {pages} {043706} (\bibinfo {year}
  {2014})}\BibitemShut {NoStop}%
\bibitem [{\citenamefont {Galvis}\ \emph {et~al.}(2015)\citenamefont {Galvis},
  \citenamefont {Herrera}, \citenamefont {Guillam{\'o}n}, \citenamefont
  {Azpeitia}, \citenamefont {Luccas}, \citenamefont {Munuera}, \citenamefont
  {Cuenca}, \citenamefont {Higuera}, \citenamefont {D{\'\i}az}, \citenamefont
  {Pazos} \emph {et~al.}}]{galvis2015three}%
  \BibitemOpen
  \bibfield  {author} {\bibinfo {author} {\bibfnamefont {J.}~\bibnamefont
  {Galvis}}, \bibinfo {author} {\bibfnamefont {E.}~\bibnamefont {Herrera}},
  \bibinfo {author} {\bibfnamefont {I.}~\bibnamefont {Guillam{\'o}n}}, \bibinfo
  {author} {\bibfnamefont {J.}~\bibnamefont {Azpeitia}}, \bibinfo {author}
  {\bibfnamefont {R.}~\bibnamefont {Luccas}}, \bibinfo {author} {\bibfnamefont
  {C.}~\bibnamefont {Munuera}}, \bibinfo {author} {\bibfnamefont
  {M.}~\bibnamefont {Cuenca}}, \bibinfo {author} {\bibfnamefont
  {J.}~\bibnamefont {Higuera}}, \bibinfo {author} {\bibfnamefont
  {N.}~\bibnamefont {D{\'\i}az}}, \bibinfo {author} {\bibfnamefont
  {M.}~\bibnamefont {Pazos}},  \emph {et~al.},\ }\href@noop {} {\bibfield
  {journal} {\bibinfo  {journal} {Review of Scientific Instruments}\ }\textbf
  {\bibinfo {volume} {86}},\ \bibinfo {pages} {013706} (\bibinfo {year}
  {2015})}\BibitemShut {NoStop}%
\bibitem [{\citenamefont {Singh}\ \emph {et~al.}(2013)\citenamefont {Singh},
  \citenamefont {Enayat}, \citenamefont {White},\ and\ \citenamefont
  {Wahl}}]{singh2013construction}%
  \BibitemOpen
  \bibfield  {author} {\bibinfo {author} {\bibfnamefont {U.~R.}\ \bibnamefont
  {Singh}}, \bibinfo {author} {\bibfnamefont {M.}~\bibnamefont {Enayat}},
  \bibinfo {author} {\bibfnamefont {S.~C.}\ \bibnamefont {White}}, \ and\
  \bibinfo {author} {\bibfnamefont {P.}~\bibnamefont {Wahl}},\ }\href@noop {}
  {\bibfield  {journal} {\bibinfo  {journal} {Review of Scientific
  Instruments}\ }\textbf {\bibinfo {volume} {84}},\ \bibinfo {pages} {013708}
  (\bibinfo {year} {2013})}\BibitemShut {NoStop}%
\bibitem [{\citenamefont {Song}\ \emph {et~al.}(2010)\citenamefont {Song},
  \citenamefont {Otte}, \citenamefont {Shvarts}, \citenamefont {Zhao},
  \citenamefont {Kuk}, \citenamefont {Blankenship}, \citenamefont {Band},
  \citenamefont {Hess},\ and\ \citenamefont {Stroscio}}]{song2010invited}%
  \BibitemOpen
  \bibfield  {author} {\bibinfo {author} {\bibfnamefont {Y.~J.}\ \bibnamefont
  {Song}}, \bibinfo {author} {\bibfnamefont {A.~F.}\ \bibnamefont {Otte}},
  \bibinfo {author} {\bibfnamefont {V.}~\bibnamefont {Shvarts}}, \bibinfo
  {author} {\bibfnamefont {Z.}~\bibnamefont {Zhao}}, \bibinfo {author}
  {\bibfnamefont {Y.}~\bibnamefont {Kuk}}, \bibinfo {author} {\bibfnamefont
  {S.~R.}\ \bibnamefont {Blankenship}}, \bibinfo {author} {\bibfnamefont
  {A.}~\bibnamefont {Band}}, \bibinfo {author} {\bibfnamefont {F.~M.}\
  \bibnamefont {Hess}}, \ and\ \bibinfo {author} {\bibfnamefont {J.~A.}\
  \bibnamefont {Stroscio}},\ }\href@noop {} {\bibfield  {journal} {\bibinfo
  {journal} {Review of Scientific Instruments}\ }\textbf {\bibinfo {volume}
  {81}},\ \bibinfo {pages} {121101} (\bibinfo {year} {2010})}\BibitemShut
  {NoStop}%
\bibitem [{\citenamefont {Misra}\ \emph {et~al.}(2013)\citenamefont {Misra},
  \citenamefont {Zhou}, \citenamefont {Drozdov}, \citenamefont {Seo},
  \citenamefont {Urban}, \citenamefont {Gyenis}, \citenamefont {Kingsley},
  \citenamefont {Jones},\ and\ \citenamefont {Yazdani}}]{Yazdani2013rsi}%
  \BibitemOpen
  \bibfield  {author} {\bibinfo {author} {\bibfnamefont {S.}~\bibnamefont
  {Misra}}, \bibinfo {author} {\bibfnamefont {B.~B.}\ \bibnamefont {Zhou}},
  \bibinfo {author} {\bibfnamefont {I.~K.}\ \bibnamefont {Drozdov}}, \bibinfo
  {author} {\bibfnamefont {J.}~\bibnamefont {Seo}}, \bibinfo {author}
  {\bibfnamefont {L.}~\bibnamefont {Urban}}, \bibinfo {author} {\bibfnamefont
  {A.}~\bibnamefont {Gyenis}}, \bibinfo {author} {\bibfnamefont {S.~C.~J.}\
  \bibnamefont {Kingsley}}, \bibinfo {author} {\bibfnamefont {H.}~\bibnamefont
  {Jones}}, \ and\ \bibinfo {author} {\bibfnamefont {A.}~\bibnamefont
  {Yazdani}},\ }\href@noop {} {\bibfield  {journal} {\bibinfo  {journal}
  {Review of Scientific Instruments}\ }\textbf {\bibinfo {volume} {84}},\
  \bibinfo {pages} {103903} (\bibinfo {year} {2013})}\BibitemShut {NoStop}%
\bibitem [{\citenamefont {Zhang}\ \emph {et~al.}(2011)\citenamefont {Zhang},
  \citenamefont {Miyamachi}, \citenamefont {Tomani{\'c}}, \citenamefont
  {Dehm},\ and\ \citenamefont {Wulfhekel}}]{Wulfhekel2011rsi}%
  \BibitemOpen
  \bibfield  {author} {\bibinfo {author} {\bibfnamefont {L.}~\bibnamefont
  {Zhang}}, \bibinfo {author} {\bibfnamefont {T.}~\bibnamefont {Miyamachi}},
  \bibinfo {author} {\bibfnamefont {T.}~\bibnamefont {Tomani{\'c}}}, \bibinfo
  {author} {\bibfnamefont {R.}~\bibnamefont {Dehm}}, \ and\ \bibinfo {author}
  {\bibfnamefont {W.}~\bibnamefont {Wulfhekel}},\ }\href@noop {} {\bibfield
  {journal} {\bibinfo  {journal} {Review of Scientific Instruments}\ }\textbf
  {\bibinfo {volume} {82}},\ \bibinfo {pages} {103702} (\bibinfo {year}
  {2011})}\BibitemShut {NoStop}%
\bibitem [{\citenamefont {Liebmann}\ \emph {et~al.}(2017)\citenamefont
  {Liebmann}, \citenamefont {Bindel}, \citenamefont {Pezzotta}, \citenamefont
  {Becker}, \citenamefont {Muckel}, \citenamefont {Johnsen}, \citenamefont
  {Saunus}, \citenamefont {Ast},\ and\ \citenamefont
  {Morgenstern}}]{Morgenstern2017rsi}%
  \BibitemOpen
  \bibfield  {author} {\bibinfo {author} {\bibfnamefont {M.}~\bibnamefont
  {Liebmann}}, \bibinfo {author} {\bibfnamefont {J.~R.}\ \bibnamefont
  {Bindel}}, \bibinfo {author} {\bibfnamefont {M.}~\bibnamefont {Pezzotta}},
  \bibinfo {author} {\bibfnamefont {S.}~\bibnamefont {Becker}}, \bibinfo
  {author} {\bibfnamefont {F.}~\bibnamefont {Muckel}}, \bibinfo {author}
  {\bibfnamefont {T.}~\bibnamefont {Johnsen}}, \bibinfo {author} {\bibfnamefont
  {C.}~\bibnamefont {Saunus}}, \bibinfo {author} {\bibfnamefont {C.~R.}\
  \bibnamefont {Ast}}, \ and\ \bibinfo {author} {\bibfnamefont
  {M.}~\bibnamefont {Morgenstern}},\ }\href@noop {} {\bibfield  {journal}
  {\bibinfo  {journal} {Review of Scientific Instruments}\ }\textbf {\bibinfo
  {volume} {88}},\ \bibinfo {pages} {123707} (\bibinfo {year}
  {2017})}\BibitemShut {NoStop}%
\bibitem [{uni()}]{unisoku}%
  \BibitemOpen
  \href@noop {} {}\bibinfo {howpublished} {Unisoku Co., Ltd., 2-4-3, Kasugano,
  Hirakata, Osaka 573-0131, Japan}\BibitemShut {NoStop}%
\bibitem [{spe()}]{specs}%
  \BibitemOpen
  \href@noop {} {}\bibinfo {howpublished} {Specs surface nano analysis GmbH,
  Voltastrasse 5, 13355 Berlin, Germany}\BibitemShut {NoStop}%
\bibitem [{\citenamefont {Scalapino}(2012)}]{Scalapino2012}%
  \BibitemOpen
  \bibfield  {author} {\bibinfo {author} {\bibfnamefont {D.~J.}\ \bibnamefont
  {Scalapino}},\ }\href {\doibase 10.1103/RevModPhys.84.1383} {\bibfield
  {journal} {\bibinfo  {journal} {Rev. Mod. Phys.}\ }\textbf {\bibinfo {volume}
  {84}},\ \bibinfo {pages} {1383} (\bibinfo {year} {2012})}\BibitemShut
  {NoStop}%
\bibitem [{\citenamefont {Davis}\ and\ \citenamefont {Lee}(2013)}]{Davis2013}%
  \BibitemOpen
  \bibfield  {author} {\bibinfo {author} {\bibfnamefont {J.~C.~S.}\
  \bibnamefont {Davis}}\ and\ \bibinfo {author} {\bibfnamefont {D.-H.}\
  \bibnamefont {Lee}},\ }\href {\doibase 10.1073/pnas.1316512110} {\bibfield
  {journal} {\bibinfo  {journal} {Proceedings of the National Academy of
  Sciences}\ }\textbf {\bibinfo {volume} {110}},\ \bibinfo {pages} {17623}
  (\bibinfo {year} {2013})}\BibitemShut {NoStop}%
\bibitem [{\citenamefont {Jourdan}, \citenamefont {Huth},\ and\ \citenamefont
  {Adrian}(1999)}]{jourdan1999superconductivity}%
  \BibitemOpen
  \bibfield  {author} {\bibinfo {author} {\bibfnamefont {M.}~\bibnamefont
  {Jourdan}}, \bibinfo {author} {\bibfnamefont {M.}~\bibnamefont {Huth}}, \
  and\ \bibinfo {author} {\bibfnamefont {H.}~\bibnamefont {Adrian}},\
  }\href@noop {} {\bibfield  {journal} {\bibinfo  {journal} {Nature}\ }\textbf
  {\bibinfo {volume} {398}},\ \bibinfo {pages} {47} (\bibinfo {year}
  {1999})}\BibitemShut {NoStop}%
\bibitem [{\citenamefont {Ishida}\ \emph {et~al.}(1998)\citenamefont {Ishida},
  \citenamefont {Mukuda}, \citenamefont {Kitaoka}, \citenamefont {Asayama},
  \citenamefont {Mao}, \citenamefont {Mori},\ and\ \citenamefont
  {Maeno}}]{ishida1998spin}%
  \BibitemOpen
  \bibfield  {author} {\bibinfo {author} {\bibfnamefont {K.}~\bibnamefont
  {Ishida}}, \bibinfo {author} {\bibfnamefont {H.}~\bibnamefont {Mukuda}},
  \bibinfo {author} {\bibfnamefont {Y.}~\bibnamefont {Kitaoka}}, \bibinfo
  {author} {\bibfnamefont {K.}~\bibnamefont {Asayama}}, \bibinfo {author}
  {\bibfnamefont {Z.}~\bibnamefont {Mao}}, \bibinfo {author} {\bibfnamefont
  {Y.}~\bibnamefont {Mori}}, \ and\ \bibinfo {author} {\bibfnamefont
  {Y.}~\bibnamefont {Maeno}},\ }\href@noop {} {\bibfield  {journal} {\bibinfo
  {journal} {Nature}\ }\textbf {\bibinfo {volume} {396}},\ \bibinfo {pages}
  {658} (\bibinfo {year} {1998})}\BibitemShut {NoStop}%
\bibitem [{\citenamefont {Tsui}, \citenamefont {Stormer},\ and\ \citenamefont
  {Gossard}(1982)}]{tsui1982two}%
  \BibitemOpen
  \bibfield  {author} {\bibinfo {author} {\bibfnamefont {D.~C.}\ \bibnamefont
  {Tsui}}, \bibinfo {author} {\bibfnamefont {H.~L.}\ \bibnamefont {Stormer}}, \
  and\ \bibinfo {author} {\bibfnamefont {A.~C.}\ \bibnamefont {Gossard}},\
  }\href@noop {} {\bibfield  {journal} {\bibinfo  {journal} {Physical Review
  Letters}\ }\textbf {\bibinfo {volume} {48}},\ \bibinfo {pages} {1559}
  (\bibinfo {year} {1982})}\BibitemShut {NoStop}%
\bibitem [{\citenamefont {Leonov}\ \emph {et~al.}(2016)\citenamefont {Leonov},
  \citenamefont {Monchesky}, \citenamefont {Romming}, \citenamefont {Kubetzka},
  \citenamefont {Bogdanov},\ and\ \citenamefont
  {Wiesendanger}}]{1367-2630-18-6-065003}%
  \BibitemOpen
  \bibfield  {author} {\bibinfo {author} {\bibfnamefont {A.~O.}\ \bibnamefont
  {Leonov}}, \bibinfo {author} {\bibfnamefont {T.~L.}\ \bibnamefont
  {Monchesky}}, \bibinfo {author} {\bibfnamefont {N.}~\bibnamefont {Romming}},
  \bibinfo {author} {\bibfnamefont {A.}~\bibnamefont {Kubetzka}}, \bibinfo
  {author} {\bibfnamefont {A.~N.}\ \bibnamefont {Bogdanov}}, \ and\ \bibinfo
  {author} {\bibfnamefont {R.}~\bibnamefont {Wiesendanger}},\ }\href
  {http://stacks.iop.org/1367-2630/18/i=6/a=065003} {\bibfield  {journal}
  {\bibinfo  {journal} {New Journal of Physics}\ }\textbf {\bibinfo {volume}
  {18}},\ \bibinfo {pages} {065003} (\bibinfo {year} {2016})}\BibitemShut
  {NoStop}%
\bibitem [{\citenamefont {Feldman}\ \emph {et~al.}(2016)\citenamefont
  {Feldman}, \citenamefont {Randeria}, \citenamefont {Gyenis}, \citenamefont
  {Wu}, \citenamefont {Ji}, \citenamefont {Cava}, \citenamefont {MacDonald},\
  and\ \citenamefont {Yazdani}}]{Feldman316}%
  \BibitemOpen
  \bibfield  {author} {\bibinfo {author} {\bibfnamefont {B.~E.}\ \bibnamefont
  {Feldman}}, \bibinfo {author} {\bibfnamefont {M.~T.}\ \bibnamefont
  {Randeria}}, \bibinfo {author} {\bibfnamefont {A.}~\bibnamefont {Gyenis}},
  \bibinfo {author} {\bibfnamefont {F.}~\bibnamefont {Wu}}, \bibinfo {author}
  {\bibfnamefont {H.}~\bibnamefont {Ji}}, \bibinfo {author} {\bibfnamefont
  {R.~J.}\ \bibnamefont {Cava}}, \bibinfo {author} {\bibfnamefont {A.~H.}\
  \bibnamefont {MacDonald}}, \ and\ \bibinfo {author} {\bibfnamefont
  {A.}~\bibnamefont {Yazdani}},\ }\href {\doibase 10.1126/science.aag1715}
  {\bibfield  {journal} {\bibinfo  {journal} {Science}\ }\textbf {\bibinfo
  {volume} {354}},\ \bibinfo {pages} {316} (\bibinfo {year}
  {2016})}\BibitemShut {NoStop}%
\bibitem [{\citenamefont {Burkov}\ and\ \citenamefont
  {Balents}(2011)}]{PhysRevLett.107.127205}%
  \BibitemOpen
  \bibfield  {author} {\bibinfo {author} {\bibfnamefont {A.~A.}\ \bibnamefont
  {Burkov}}\ and\ \bibinfo {author} {\bibfnamefont {L.}~\bibnamefont
  {Balents}},\ }\href {\doibase 10.1103/PhysRevLett.107.127205} {\bibfield
  {journal} {\bibinfo  {journal} {Phys. Rev. Lett.}\ }\textbf {\bibinfo
  {volume} {107}},\ \bibinfo {pages} {127205} (\bibinfo {year}
  {2011})}\BibitemShut {NoStop}%
\bibitem [{\citenamefont {Baumann}(2011)}]{Baumann2011}%
  \BibitemOpen
  \bibfield  {author} {\bibinfo {author} {\bibfnamefont {D.}~\bibnamefont
  {Baumann}},\ }\href@noop {} {\emph {\bibinfo {title} {Aufbau eines
  ultrahochaufl{\"o}senden Tieftemperatur-Rastertunnelmikroskops}}}\ (\bibinfo
  {publisher} {mbvberlin},\ \bibinfo {year} {2011})\ \bibinfo {note} {{ISBN:}
  978-3-86387-137-6}\BibitemShut {NoStop}%
\bibitem [{\citenamefont {Salazar}(2016)}]{Salazar2016}%
  \BibitemOpen
  \bibfield  {author} {\bibinfo {author} {\bibfnamefont {C.}~\bibnamefont
  {Salazar}},\ }\emph {\bibinfo {title} {Scanning tunneling microscopy on low
  dimensional systems: dinickel molecular complexes and iron nanostructures}},\
  \href
  {http://www.qucosa.de/recherche/frontdoor/cache.off?tx_slubopus4frontend%5Bid%5D=21157}
  {Ph.D. thesis},\ \bibinfo  {school} {Technische Universit{\"a}t Dresden}
  (\bibinfo {year} {2016}),\ \bibinfo {note} {{URL:}
  http://nbn-resolving.de/urn:nbn:de:bsz:14-qucosa-211572}\BibitemShut
  {NoStop}%
\bibitem [{\citenamefont {Scheffler}(2015)}]{Scheffler2015}%
  \BibitemOpen
  \bibfield  {author} {\bibinfo {author} {\bibfnamefont {M.}~\bibnamefont
  {Scheffler}},\ }\emph {\bibinfo {title} {Microscopic tunneling experiments on
  atomic impurities in graphene and on magnetic thin films}},\ \href@noop {}
  {Ph.D. thesis},\ \bibinfo  {school} {Technische Universit{\"a}t Dresden}
  (\bibinfo {year} {2015}),\ \bibinfo {note}
  {{URL:}http://nbn-resolving.de/urn:nbn:de:bsz:14-qucosa-174831}\BibitemShut
  {NoStop}%
\bibitem [{bil()}]{bilz}%
  \BibitemOpen
  \href@noop {} {}\bibinfo {howpublished} {Bilz Vibration Technology AG,
  B{\"o}blinger Strasse 25, D-71229 Leonberg, Germany}\BibitemShut {NoStop}%
\bibitem [{jan()}]{janis}%
  \BibitemOpen
  \href@noop {} {}\bibinfo {howpublished} {Janis Research Company Inc., 2 Jewel
  Drive, Wilmington, MA 01887-3350 USA}\BibitemShut {NoStop}%
\bibitem [{\citenamefont {Raccanelli}, \citenamefont {Reichertz},\ and\
  \citenamefont {Kreysa}(2001)}]{Raccanelli2001}%
  \BibitemOpen
  \bibfield  {author} {\bibinfo {author} {\bibfnamefont {A.}~\bibnamefont
  {Raccanelli}}, \bibinfo {author} {\bibfnamefont {L.~A.}\ \bibnamefont
  {Reichertz}}, \ and\ \bibinfo {author} {\bibfnamefont {E.}~\bibnamefont
  {Kreysa}},\ }\href@noop {} {\bibfield  {journal} {\bibinfo  {journal}
  {Cryogenics}\ }\textbf {\bibinfo {volume} {41}},\ \bibinfo {pages} {763}
  (\bibinfo {year} {2001})}\BibitemShut {NoStop}%
\bibitem [{lak()}]{lakeshore}%
  \BibitemOpen
  \href@noop {} {}\bibinfo {howpublished} {Lake Shore Cryotronics Inc., 575
  McCorkle Blvd, Westerville, OH 43082-8699, USA}\BibitemShut {NoStop}%
\bibitem [{\citenamefont {Haude}(2001)}]{haude2001}%
  \BibitemOpen
  \bibfield  {author} {\bibinfo {author} {\bibfnamefont {D.}~\bibnamefont
  {Haude}},\ }\emph {\bibinfo {title} {Rastertunnelspektroskopie auf der
  InAs(110) Oberfläche:Untersuchungen an drei-, zwei-, und nulldimensionalen
  Elektronensystemen im Magnetfeld}},\ \href@noop {} {Ph.D. thesis},\ \bibinfo
  {school} {Hamburg University} (\bibinfo {year} {2001})\BibitemShut {NoStop}%
\bibitem [{\citenamefont {Wachowiak}(2003)}]{wachowiak2003}%
  \BibitemOpen
  \bibfield  {author} {\bibinfo {author} {\bibfnamefont {A.}~\bibnamefont
  {Wachowiak}},\ }\emph {\bibinfo {title} {Aufbau einer
  300mK-Ultrahochvakuum-Rastertunnelmikroskopie-Anlage mit 14 Tesla Magnet und
  spin-polarisierte Rastertunnelspektroskopie an ferromagnetischen
  Fe-Inseln}},\ \href@noop {} {Ph.D. thesis},\ \bibinfo  {school} {Hamburg
  University} (\bibinfo {year} {2003})\BibitemShut {NoStop}%
\bibitem [{\citenamefont {H{\"a}nke}(2005)}]{hanke2005}%
  \BibitemOpen
  \bibfield  {author} {\bibinfo {author} {\bibfnamefont {T.}~\bibnamefont
  {H{\"a}nke}},\ }\emph {\bibinfo {title} {A new variable-temperature scanning
  tunneling microscope and temperature-dependent spin-polarized scanning
  tunneling spectroscopy on the Cr(001) surface}},\ \href@noop {} {Ph.D.
  thesis},\ \bibinfo  {school} {Hamburg University} (\bibinfo {year}
  {2005})\BibitemShut {NoStop}%
\bibitem [{\citenamefont {Schlegel}\ \emph {et~al.}(2014)\citenamefont
  {Schlegel}, \citenamefont {H{\"a}nke}, \citenamefont {Baumann}, \citenamefont
  {Kaiser}, \citenamefont {Nag}, \citenamefont {Voigtl{\"a}nder}, \citenamefont
  {Lindackers}, \citenamefont {B{\"u}chner},\ and\ \citenamefont
  {Hess}}]{Schlegel2014}%
  \BibitemOpen
  \bibfield  {author} {\bibinfo {author} {\bibfnamefont {R.}~\bibnamefont
  {Schlegel}}, \bibinfo {author} {\bibfnamefont {T.}~\bibnamefont {H{\"a}nke}},
  \bibinfo {author} {\bibfnamefont {D.}~\bibnamefont {Baumann}}, \bibinfo
  {author} {\bibfnamefont {M.}~\bibnamefont {Kaiser}}, \bibinfo {author}
  {\bibfnamefont {P.~K.}\ \bibnamefont {Nag}}, \bibinfo {author} {\bibfnamefont
  {R.}~\bibnamefont {Voigtl{\"a}nder}}, \bibinfo {author} {\bibfnamefont
  {D.}~\bibnamefont {Lindackers}}, \bibinfo {author} {\bibfnamefont
  {B.}~\bibnamefont {B{\"u}chner}}, \ and\ \bibinfo {author} {\bibfnamefont
  {C.}~\bibnamefont {Hess}},\ }\href {\doibase 10.1063/1.4862817} {\bibfield
  {journal} {\bibinfo  {journal} {Review of Scientific Instruments}\ }\textbf
  {\bibinfo {volume} {85}},\ \bibinfo {pages} {013706} (\bibinfo {year}
  {2014})}\BibitemShut {NoStop}%
\bibitem [{pat()}]{patent}%
  \BibitemOpen
  \href@noop {} {}\bibinfo {howpublished} {S. H. Pan, Piezo-electric Motor,
  International Patent Publication No. WO 93/19494, World Intellectual Property
  Organization, 1993}\BibitemShut {NoStop}%
\bibitem [{epo()}]{epotek}%
  \BibitemOpen
  \href@noop {} {}\bibinfo {howpublished} {Epoxy Technology Inc., 14 Fortune
  Drive, Billerica, MA 01821, USA}\BibitemShut {NoStop}%
\bibitem [{\citenamefont {Binnig}\ and\ \citenamefont
  {Smith}(1986)}]{binnig1986}%
  \BibitemOpen
  \bibfield  {author} {\bibinfo {author} {\bibfnamefont {G.}~\bibnamefont
  {Binnig}}\ and\ \bibinfo {author} {\bibfnamefont {D.}~\bibnamefont {Smith}},\
  }\href@noop {} {\bibfield  {journal} {\bibinfo  {journal} {Review of
  Scientific Instruments}\ }\textbf {\bibinfo {volume} {57}},\ \bibinfo {pages}
  {1688} (\bibinfo {year} {1986})}\BibitemShut {NoStop}%
\bibitem [{dat()}]{dataforth}%
  \BibitemOpen
  \href@noop {} {}\bibinfo {howpublished} {Dataforth Corporation, 3331 E.
  Hemisphere Loop Tucson, AZ 85706-5011, USA}\BibitemShut {NoStop}%
\bibitem [{fem()}]{femto}%
  \BibitemOpen
  \href@noop {} {}\bibinfo {howpublished} {FEMTO Messtechnik GmbH,
  Klosterstrasse 64, 10179 Berlin, Germany}\BibitemShut {NoStop}%
\bibitem [{tus()}]{tusonix}%
  \BibitemOpen
  \href@noop {} {}\bibinfo {howpublished} {Tusonix Inc., 7741 N. Business Park
  Dr., Tucson, AZ 85743, USA}\BibitemShut {NoStop}%
\bibitem [{\citenamefont {Morozov}\ \emph {et~al.}(2010)\citenamefont
  {Morozov}, \citenamefont {Boltalin}, \citenamefont {Volkova}, \citenamefont
  {Vasiliev}, \citenamefont {Kataeva}, \citenamefont {Stockert}, \citenamefont
  {Abdel-Hafiez}, \citenamefont {Bombor}, \citenamefont {Bachmann},
  \citenamefont {Harnagea} \emph {et~al.}}]{morozov2010single}%
  \BibitemOpen
  \bibfield  {author} {\bibinfo {author} {\bibfnamefont {I.}~\bibnamefont
  {Morozov}}, \bibinfo {author} {\bibfnamefont {A.}~\bibnamefont {Boltalin}},
  \bibinfo {author} {\bibfnamefont {O.}~\bibnamefont {Volkova}}, \bibinfo
  {author} {\bibfnamefont {A.}~\bibnamefont {Vasiliev}}, \bibinfo {author}
  {\bibfnamefont {O.}~\bibnamefont {Kataeva}}, \bibinfo {author} {\bibfnamefont
  {U.}~\bibnamefont {Stockert}}, \bibinfo {author} {\bibfnamefont
  {M.}~\bibnamefont {Abdel-Hafiez}}, \bibinfo {author} {\bibfnamefont
  {D.}~\bibnamefont {Bombor}}, \bibinfo {author} {\bibfnamefont
  {A.}~\bibnamefont {Bachmann}}, \bibinfo {author} {\bibfnamefont
  {L.}~\bibnamefont {Harnagea}},  \emph {et~al.},\ }\href@noop {} {\bibfield
  {journal} {\bibinfo  {journal} {Crystal Growth \& Design}\ }\textbf {\bibinfo
  {volume} {10}},\ \bibinfo {pages} {4428} (\bibinfo {year}
  {2010})}\BibitemShut {NoStop}%
\bibitem [{\citenamefont {Nag}\ \emph {et~al.}(2016)\citenamefont {Nag},
  \citenamefont {Schlegel}, \citenamefont {Baumann}, \citenamefont {Grafe},
  \citenamefont {Beck}, \citenamefont {Wurmehl}, \citenamefont {B{\"u}chner},\
  and\ \citenamefont {Hess}}]{nag2016two}%
  \BibitemOpen
  \bibfield  {author} {\bibinfo {author} {\bibfnamefont {P.}~\bibnamefont
  {Nag}}, \bibinfo {author} {\bibfnamefont {R.}~\bibnamefont {Schlegel}},
  \bibinfo {author} {\bibfnamefont {D.}~\bibnamefont {Baumann}}, \bibinfo
  {author} {\bibfnamefont {H.-J.}\ \bibnamefont {Grafe}}, \bibinfo {author}
  {\bibfnamefont {R.}~\bibnamefont {Beck}}, \bibinfo {author} {\bibfnamefont
  {S.}~\bibnamefont {Wurmehl}}, \bibinfo {author} {\bibfnamefont
  {B.}~\bibnamefont {B{\"u}chner}}, \ and\ \bibinfo {author} {\bibfnamefont
  {C.}~\bibnamefont {Hess}},\ }\href@noop {} {\bibfield  {journal} {\bibinfo
  {journal} {Scientific reports}\ }\textbf {\bibinfo {volume} {6}} (\bibinfo
  {year} {2016})}\BibitemShut {NoStop}%
\bibitem [{\citenamefont {Schlegel}\ \emph {et~al.}(2016)\citenamefont
  {Schlegel}, \citenamefont {Nag}, \citenamefont {Baumann}, \citenamefont
  {Beck}, \citenamefont {Wurmehl}, \citenamefont {B{\"u}chner},\ and\
  \citenamefont {Hess}}]{schlegel2016defect}%
  \BibitemOpen
  \bibfield  {author} {\bibinfo {author} {\bibfnamefont {R.}~\bibnamefont
  {Schlegel}}, \bibinfo {author} {\bibfnamefont {P.}~\bibnamefont {Nag}},
  \bibinfo {author} {\bibfnamefont {D.}~\bibnamefont {Baumann}}, \bibinfo
  {author} {\bibfnamefont {R.}~\bibnamefont {Beck}}, \bibinfo {author}
  {\bibfnamefont {S.}~\bibnamefont {Wurmehl}}, \bibinfo {author} {\bibfnamefont
  {B.}~\bibnamefont {B{\"u}chner}}, \ and\ \bibinfo {author} {\bibfnamefont
  {C.}~\bibnamefont {Hess}},\ }\href@noop {} {\bibfield  {journal} {\bibinfo
  {journal} {physica status solidi (b)}\ } (\bibinfo {year}
  {2016})}\BibitemShut {NoStop}%
\bibitem [{\citenamefont {Bode}\ \emph {et~al.}(2007)\citenamefont {Bode},
  \citenamefont {Krause}, \citenamefont {Berbil-Bautista}, \citenamefont
  {Heinze},\ and\ \citenamefont {Wiesendanger}}]{bode2007preparation}%
  \BibitemOpen
  \bibfield  {author} {\bibinfo {author} {\bibfnamefont {M.}~\bibnamefont
  {Bode}}, \bibinfo {author} {\bibfnamefont {S.}~\bibnamefont {Krause}},
  \bibinfo {author} {\bibfnamefont {L.}~\bibnamefont {Berbil-Bautista}},
  \bibinfo {author} {\bibfnamefont {S.}~\bibnamefont {Heinze}}, \ and\ \bibinfo
  {author} {\bibfnamefont {R.}~\bibnamefont {Wiesendanger}},\ }\href@noop {}
  {\bibfield  {journal} {\bibinfo  {journal} {Surface science}\ }\textbf
  {\bibinfo {volume} {601}},\ \bibinfo {pages} {3308} (\bibinfo {year}
  {2007})}\BibitemShut {NoStop}%
\bibitem [{\citenamefont {Bardeen}, \citenamefont {Cooper},\ and\ \citenamefont
  {Schrieffer}(1957)}]{PhysRev.106.162}%
  \BibitemOpen
  \bibfield  {author} {\bibinfo {author} {\bibfnamefont {J.}~\bibnamefont
  {Bardeen}}, \bibinfo {author} {\bibfnamefont {L.~N.}\ \bibnamefont {Cooper}},
  \ and\ \bibinfo {author} {\bibfnamefont {J.~R.}\ \bibnamefont {Schrieffer}},\
  }\href {\doibase 10.1103/PhysRev.106.162} {\bibfield  {journal} {\bibinfo
  {journal} {Phys. Rev.}\ }\textbf {\bibinfo {volume} {106}},\ \bibinfo {pages}
  {162} (\bibinfo {year} {1957})}\BibitemShut {NoStop}%
\bibitem [{\citenamefont {Tinkham}(1996)}]{Tinkham}%
  \BibitemOpen
  \bibfield  {author} {\bibinfo {author} {\bibfnamefont {M.}~\bibnamefont
  {Tinkham}},\ }\href@noop {} {\emph {\bibinfo {title} {Introduction to
  superconductivity}}},\ \bibinfo {edition} {2nd}\ ed.\ (\bibinfo  {publisher}
  {McGraw-Hill Inc.},\ \bibinfo {year} {1996})\BibitemShut {NoStop}%
\bibitem [{\citenamefont {Pan}, \citenamefont {Hudson},\ and\ \citenamefont
  {Davis}(1998)}]{pan1998vacuum}%
  \BibitemOpen
  \bibfield  {author} {\bibinfo {author} {\bibfnamefont {S.}~\bibnamefont
  {Pan}}, \bibinfo {author} {\bibfnamefont {E.}~\bibnamefont {Hudson}}, \ and\
  \bibinfo {author} {\bibfnamefont {J.}~\bibnamefont {Davis}},\ }\href@noop {}
  {\bibfield  {journal} {\bibinfo  {journal} {Applied physics letters}\
  }\textbf {\bibinfo {volume} {73}},\ \bibinfo {pages} {2992} (\bibinfo {year}
  {1998})}\BibitemShut {NoStop}%
\bibitem [{\citenamefont {Townsend}\ and\ \citenamefont
  {Sutton}(1962)}]{townsend1962investigation}%
  \BibitemOpen
  \bibfield  {author} {\bibinfo {author} {\bibfnamefont {P.}~\bibnamefont
  {Townsend}}\ and\ \bibinfo {author} {\bibfnamefont {J.}~\bibnamefont
  {Sutton}},\ }\href@noop {} {\bibfield  {journal} {\bibinfo  {journal}
  {Physical Review}\ }\textbf {\bibinfo {volume} {128}},\ \bibinfo {pages}
  {591} (\bibinfo {year} {1962})}\BibitemShut {NoStop}%
\bibitem [{\citenamefont {von Bergmann}(2004)}]{Kirsten2004}%
  \BibitemOpen
  \bibfield  {author} {\bibinfo {author} {\bibfnamefont {K.}~\bibnamefont {von
  Bergmann}},\ }\emph {\bibinfo {title} {Iron nanostructures studied by
  spin-polarised scanning tunneling spectroscopy}},\ \href@noop {} {Ph.D.
  thesis},\ \bibinfo  {school} {Hamburg University} (\bibinfo {year}
  {2004})\BibitemShut {NoStop}%
\bibitem [{\citenamefont {Kubetzka}\ \emph {et~al.}(2001)\citenamefont
  {Kubetzka}, \citenamefont {Pietzsch}, \citenamefont {Bode},\ and\
  \citenamefont {Wiesendanger}}]{kubetzka2001magnetism}%
  \BibitemOpen
  \bibfield  {author} {\bibinfo {author} {\bibfnamefont {A.}~\bibnamefont
  {Kubetzka}}, \bibinfo {author} {\bibfnamefont {O.}~\bibnamefont {Pietzsch}},
  \bibinfo {author} {\bibfnamefont {M.}~\bibnamefont {Bode}}, \ and\ \bibinfo
  {author} {\bibfnamefont {R.}~\bibnamefont {Wiesendanger}},\ }\href@noop {}
  {\bibfield  {journal} {\bibinfo  {journal} {Physical Review B}\ }\textbf
  {\bibinfo {volume} {63}},\ \bibinfo {pages} {140407} (\bibinfo {year}
  {2001})}\BibitemShut {NoStop}%
\end{thebibliography}
%

\end{document}